\documentclass[10pt,a4paper]{iopart}

\usepackage{rotating,graphicx}
\usepackage{amsfonts}%
\usepackage{amssymb}%
\usepackage{caption}
\usepackage[utf8]{inputenc}
\usepackage[english]{babel} 
\usepackage[T1]{fontenc}
\usepackage{color}
\usepackage[colorlinks]{hyperref}
\usepackage{setspace}
\usepackage{layouts}
\begin{document}

\title[An experimental characterization of core turbulence regimes in W7-X]{An experimental characterization of core turbulence regimes in Wendelstein 7-X}
\author{D. Carralero$^1$, T. Estrada$^1$, E. Maragkoudakis$^1$, T. Windisch$^2$, J. A. Alonso$^1$, M. Beurskens$^2$, S. Bozhenkov$^2$, I. Calvo$^1$, H. Damm$^2$, O. Ford$^2$, G. Fuchert$^2$, J. M. García-Regaña$^1$, N. Pablant$^3$, E. Sánchez$^1$, E. Pasch$^2$, J.L. Velasco$^1$ and the Wendelstein 7-X team.}
\address{$^1$ Laboratorio Nacional de Fusión. CIEMAT, 28040 Madrid, Spain.}
\address{$^2$ Max-Planck-Institut für Plasmaphysik, D-17491 Greifswald, Germany.}
\address{$^3$ Princeton Plasma Phys Lab, 100 Stellarator Rd, Princeton, NJ 08540 USA} 
\ead{daniel.carralero@ciemat.es}

\begin{abstract}

First results from the optimized helias Wendelstein 7-X stellarator (W7-X) have shown that core transport is no longer mostly neoclassical, as is the case in previous kinds of stellarators. Instead, power balance analysis has shown that turbulent transport poses a serious limitation to the global performance of the machine. Several studies have found this particularly relevant for ion transport, with core ion temperatures becoming clamped at relatively low values of $T_{i} \simeq 1.7$ keV, except in the few scenarios in which turbulence can be suppressed. In order to understand the precise turbulent mechanisms at play and thus design improved performance scenarios, it is important to have a clear understanding of the parametric dependencies of turbulent fluctuations, and the relation between them and turbulent transport. As a first step in this direction, in this work we use Doppler reflectometry measurements carried out during a number of relevant operational scenarios to provide a systematic characterization of ion-scale ($k_\perp\rho_i\simeq 1$) density fluctuations in the core of W7-X. Then, we study the relation between fluctuation amplitude and plasma profiles and show how distinct regimes can be defined for the former, depending on normalized gradients $a/L_{ne}$ and $a/L_{Ti}$. Furthermore, we discuss the importance of other potentially relevant parameters such as $T_e/T_i$, $E_r$ or collisionality. Comparing the different regimes, we find that turbulence amplitude depends generally on the gradient ratio $\eta_i=L_{ne}/L_{Ti}$, as would be expected for ITG modes, with the exception of a range of discharges, for which turbulence suppression may be better explained by an ITG to TEM transition triggered by a drop in collisionality. Finally, we show a number of scenarios under which $T_{i,core} > 1.7$ keV is achieved and how core fluctuations are suppressed in all of them, thus providing experimental evidence of microturbulence being the main responsible for the limited ion confinement in W7-X.
\end{abstract}

\maketitle

\section{Introduction}\label{intro}

Optimized stellarators are regarded as a potential pathway for fusion reactors in which the general advantages of stellarators over the tokamaks  are conserved (such as stationary pulses, absence of disruptions, etc.) while neoclassical fluxes are reduced to acceptable levels, thus solving or substantially improving one of the main disadvantages of the stellarator approach. The largest device of this kind is currently Wendelstein 7-X (W7-X), an optimized helias stellarator in which quasi-isodinamicity is used to reduce drift orbit losses caused by the three-dimensional magnetic field inhomogeneity combined with collisions, thus improving the traditional low confinement of stellarators at low collisionalities. This device, which began operation in 2016 \cite{Wolf17,Klinger17}, has conducted its last operational phase (OP1.2b) between the months of August and October of 2018, featuring diverted plasmas and wall boronization, and achieving 200 MJ discharges and complete detachment in the island divertor \cite{Pedersen19}. While the results of the first experimental campaign confirm that neoclassical transport in W7-X has been reduced with respect to non-optimized stellarators \cite{Beidler19}, comparison of neoclassical predictions with experimental measurements of the total energy transport suggests that under most scenarios turbulence accounts for a large fraction of the total transport, even in the core \cite{Dinklage18,Bozhenkov19}. This dominant role of turbulence has been invoked to explain the limited performance of W7-X so far, with energy confinement times, $\tau_E$, falling below the empirical ISS04 scaling for high density plasmas \cite{Fuchert20}. Perhaps, the most obvious effect of this is a cap on the core ion temperature, $T_{i,core}$, which is limited to $T_{i,core} \lesssim 1.7$ keV in gas puff discharges with electron-cyclotron resonance heating (ECRH) \cite{Bozhenkov19}. This upper limit is not affected by collisionality levels or the injected heating power, P$_{ECRH}$, at least in the available range, and is well below the values expected for neoclassical predictions \cite{Beruskens20, Beurskens21}, indicating that this limitation is related to the high turbulent transport. During the last experimental campaigns, a number of transient scenarios were found in which this clamping was relaxed and higher $T_{i,core}$ values could be observed. The most prominent of them is the "high performance" (HP) operation, equivalent to similar ones reported both in tokamaks \cite{Greenwald85, Tubbing91, Quigley04} and other stellarators \cite{Yamada00}. It consists of a phase after the injection of pellets during ECRH operation in which record values of $T_{i,core} \simeq 2.5$ keV and stored energy, $W_{dia} \simeq 1$ MJ are achieved while core transport is substantially reduced, matching neoclassical levels \cite{Bozhenkov20}. Consistently, a substantial reduction of turbulence amplitude is observed experimentally during this phase \cite{Stechow20,Estrada21}.  Besides HP operation, similar transitory phases also featuring enhanced density gradients and reduced turbulence have been observed in W7-X following the injection of impurities \cite{Lunsford21}. Finally, some scenarios in which $T_{i,core} > 1.7$ keV is temporarily achieved involve the use of neutral beam injection (NBI) to obtain a steep density gradient \cite{Ford19}. However, the physical picture is more complicated in those, as some degree of additional ECRH is required, indicating some complex interplay between transport, fueling and heating.\\  

These results indicate that core turbulent transport will be much more relevant in optimized stellarators - particularly for the ion channel-, and that a good understanding of the physical mechanisms underlying the turbulence will be critical to achieve sufficient global confinement to reach the $T_i = 5$ keV physics goal of W7-X in the road of a stellarator reactor based on the helias concept \cite{Wolf16}. In this work, we set out to approach this objective by performing a systematic characterization of experimental measurements of ion-scale core turbulence in W7-X, carried out with a Doppler reflectometer (DR) under a wide range of relevant scenarios from the last operational phase, including some of the aforementioned situations in which the  $T_{i,core} \lesssim 1.7$ keV limit is relaxed. After this, we aim to evaluate the impact of different plasma parameters on fluctuation amplitude in order to identify the underlying instabilities and pave the way for a future discussion on the mechanisms regulating turbulent transport and their impact in it the general performance of the machine. It is generally accepted that ion-scale turbulence in magnetically confined fusion plasmas is dominated by ion-temperature gradient (ITG) modes \cite{Doyle07}, with spatial scales close to the ion gyroradius, $k_\perp\rho_i \simeq 1$, where $k_\perp$ is the mode wavelength in the binormal direction. This kind of drift wave instability typically features a critical stability threshold on the gradient ratio $\eta_{i} = L_{n_i}/L_{T_i}$ \cite{Horton99}, where $L_\alpha$ is the local gradient scale of the parameter $\alpha$ (in fact, in the 80s and 90s literature, ITG modes are sometimes referred to as the "$\eta_i$-modes" \cite{Rogister88}). In scenarios with flat density profiles, $L_n \rightarrow \infty$, the ITG mode is destabilized when a certain normalized gradient $R/L_{T_i}$ is exceeded instead \cite{Tang86}. While this physical picture was initially developed for tokamaks, it is considered to be valid also for stellarators. In particular, recent linear calculations show this for W7-X, with $\eta_i$ thresholds in the range of $\eta_{i,c} \simeq 1$ \cite{Riemann15,Zocco18}. This strong dependency on plasma profiles has been used to explain the HP regime as the result of ITG mode suppression: indeed, during the pellet phase a strong density gradient is built while core temperatures drop as the result of the cold ion and electron populations resulting from the pellets. Then, after the pellet injection has substantially decreased $\eta_i$ values, the confinement is improved and turbulent transport reduced. Consistently with this interpretation, the  reduction in turbulent heat flux is only observed in the ion channel, with electron heat transport remaining clearly above neoclassical levels during the HP phase \cite{Bozhenkov20}. This relation to ITG turbulence could as well explain the transient nature of this regime: after turbulence is suppressed, $T_i$ would increase as a result of the enhanced confinement and eventually the density and its peaking would decay, recovering high  $\eta_i$ values, thus destabilizing turbulence again and ending the transitory phase.\\

Density and temperature gradients are determinant for the ITG mode, and therefore the main quantities to be compared to fluctuation measurements for their characterization. However, there are other parameters that should be considered, as they may affect the stability of these modes: first, although gradients are enough to excite an ITG mode in the slab approximation, when more realistic geometries are considered the destabilizing effect of the $\nabla B$ and curvature drifts become important. As a result, ITG modes appear preferentially in regions of bad curvature (prone to interchange instabilities) in toroidal geometries, which in the case of W7-X comprise a localized band around the outer midplane in the bean section \cite{Kornilov04,Xanthopoulos14,Banon20}. Another factor influencing ITG turbulence is the local $T_e/T_i$ ratio, which reduces the stability for a given set of gradients when $T_e/T_i >1$ (and increases it in the opposite case). This has been discussed in the literature for the case of W7-X, for which linear simulations carried out with the gyrokinetic code GENE found the ITG critical temperature gradient to be a decreasing function of $T_e/T_i$ ratio, while growth rates increased with it \cite{Zocco18,Beurskens21}. As well, there are experimental evidence of both density fluctuation amplitude and turbulent impurity transport being enhanced in W7-X by increasing  $T_e/T_i$ when laser blow-off is used to inject various impurities under different levels of $P_{ECRH}$ heating \cite{Wegner20}. Radial electric fields, $E_r$, may also have a noticeable influence in turbulent fluctuations, at least through two separate mechanisms: first, local $E_r$ induces an $\mathbf{E} \times \mathbf{B}$ rotation in the plasma which may displace fluctuations on each flux surface away from the regions of instability \cite{Riemann15}. Second, if a strong $E_r$ shear is present, the resulting shear flow may decorrelate turbulent structures, thereby reducing turbulence \cite{Bigliari90}. Turbulence suppression by $E_r$ shear has been observed in W7-X \cite{CarraleroIRW, AKF20} and the stabilization effect of $E_r$ has been studied by means of gyrokinetic simulations, which confirmed the existence of this mechanism but found its effect to be less than that of profiles \cite{Xanthopoulos20,Estrada21}. Finally, ITG modes are not the only possible instabilities in the $k_\perp\rho_i \simeq 1$ scale. Another mode which may play a role in ion heat transport is the trapped electron mode (TEM) \cite{Coppi74}, a loose term designating a family of instabilities driven by either the $n_e$ or the $T_e$ gradient. This mode, which has been found to be relevant for transport in stellarators \cite{Sanchez19}, is generated by the electron population having magnetically trapped orbits. As a result, it is most unstable in regions in which bad curvature is in the vicinity of a local minimum of $|\mathbf{B}|$ along the field line \cite{Hellander13} and only can be excited for sufficiently low values of electron collision frequency, $\nu_{e}$. TEMs may interact in complex ways with ITG modes, and a transition between them may occur if the relation between the relevant parameters ($\eta_i$, $\eta_e$, $\nu_{e}$) changes. In particular, such transition from TEM to ITG has been observed in tokamaks when collisionality was raised above a certain threshold, in good agreement with theoretical predictions by the gyrokinetic code GS2 \cite{Ryter05}. In order to provide a characterization of core turbulence, all these parameters are to be measured and discussed along with the fluctuations.\\

 Two final comments regarding the scope of the study are in order: first, only data from a single magnetic configuration (standard) will be considered, leaving the analysis of the influence of magnetic geometry parameters for future work. Second, the focus of this paper is mostly to provide an empirical description of the different regimes of turbulence in relevant scenarios of operation, and the relation between measured fluctuations and the previously discussed parameters. In this sense, it must be stressed that general performance is determined by global turbulent transport, which is measured by power balance analysis \cite{Dinklage13,Bozhenkov20}, or only recently in stellarators, calculated by means of nonlinear gyrokinetic simulations \cite{Sanchez20,Xanthopoulos20,Regana21}. The relation between DR measurements of local fluctuations and such global turbulent transport is by no means trivial and, leaving aside some preliminary results which can be advanced from the data presented here, is considered out of the scope of this work and will be addressed at length in a forthcoming one.  The rest of the paper is organized as follows: the main diagnostics and analysis methodology are described in section  \ref{Met} and the selected scenarios in section \ref{data}. Then, experimental results are presented in section \ref{exp} and compared to the various parameters in section \ref{drives}. To conclude, results are discussed and conclusions outlined in section \ref{discussion}.

\section{Methodology}\label{Met}

The main diagnostic used to characterize turbulence in this study is a Doppler reflectometer, a device capable of providing a $k_\perp$-selective measure of the amplitude of density fluctuations as well as their laboratory-frame perpendicular velocity based respectively on the power and Doppler shift of the backscattered power of a microwave beam launched into the plasma. The DR was first used to measure the radial electric field in the W7-AS stellarator \cite{Hirsch99} and the Tore Supra tokamak \cite{Zou99} and has since become a standard diagnostic for the study of $E_r$ and turbulence in both kinds of devices \cite{Happel09,Tokuzawa12,Estrada19,Happel15}, including some currently in development such as JT60-SA \cite{CarraleroJT60} or ITER \cite{Muscatello20}. A DR is based on the fact that an electromagnetic wave launched into a plasma propagates radially until a cutoff layer is reached, defined as the point in which its refractive index, $N$, achieves its minimum value. If the beam is launched with an oblique angle, and turbulence of a certain amplitude is present at the cutoff layer, a fraction of the power of the incident beam is backscattered by density fluctuations with a perpendicular wavenumber fulfilling the Bragg criterion, $k_\perp=2k_i$, where $k_i$ is the wavenumber of the incident beam \cite{Hirsch01}.  In a monostatic DR, this backscattered wave travels back to the emitting antenna, where it is measured. For low turbulence levels the backscattered power, $S$, is related to the density fluctuations through the simple relation: 

\begin{equation}
	S \propto (\delta n)^2, \label{eq1}
\end{equation}

where $\delta n$ is the amplitude of density fluctuations \cite{Gusakov04, Blanco08}. Besides, the frequency of the backscattered wave is shifted due to the Doppler effect induced by the perpendicular velocity, $u_\perp$, of the density fluctuations of $k_\perp$ wavenumber. Therefore, both $S$ and $u_\perp$ can be obtained from the measurement of the Doppler peak characteristics $\omega_D$, $A_D$, and $\Delta \omega_D$ (respectively the Doppler frequency shift, and the peak height and width): first, the backscattered power can be estimated as $S = A_D \Delta \omega_D/2\pi$. In order to determine $\delta n$ from $S$, it must be noted that both the microwave generator and the various microwave transmission systems feature a different power response depending on the beam frequency. As a result, a power calibration of the complete system is required for each frequency in order to obtain a general form of equation (\ref{eq1}). As explained in \cite{Estrada21}, this is done by a combination of a direct calibration of the ex-vessel system components, information provided by the manufacturer of the ones inside the vacuum vessel and systematic error analysis methods. Then, $\omega_D = \mathbf{u}\cdot\mathbf{k} \simeq u_\perp k_\perp$ permits the calculation of $u_\perp$ if the wavenumber is known. This rotation measurement can then be used to estimate the local radial electric field, $E_r$, by assuming that $u_\perp \simeq v_E$, where $v_E =|\mathbf{E}\times\mathbf{B}|/B^2$ is the $E\times B$ bulk velocity of the plasma. This approximation is only correct if this velocity dominates the turbulence phase velocity $v_E \gg v_{ph}$. Such approach has often found to be correct in stellarator plasmas \cite{Hirsch01,Estrada09,AKF2019} and is supported at the core of W7-X by the general good agreement between $E_r$ profiles calculated form DR data and neoclassical code predictions \cite{Carralero20,Estrada21}. A more detailed discussion about the validity of this approximation and its limits can be found elsewhere \cite{Windisch19}.\\

The DR provides a good localization of turbulence amplitude and rotation both in real space and in the wavenumber spectrum: if the wave is polarized in the O-mode (as in the DR used in this study), $N$ is only a function of the beam frequency, $\omega$, and the local electron density,

\begin{equation}
	N^2=\biggl(1-\frac{\omega_p^2}{\omega^2}\biggr), \label{eq0}
\end{equation}

where $\omega_p=(n_ee^2/\epsilon_0m_e)^{1/2}$ is the plasma frequency and $n_e$, $e$ and $m_e$ stand for the electron density, charge and mass. Under the slab approximation and assuming small refraction, the beam incidence at the cutoff layer can be associated to the launching angle of the beam, $\theta_0$. In this case, the cutoff density, $n_{co}$, can be calculated by setting the condition $N_{co}=\sin{\theta_0}$. Also, the wavenumber of the backscattering turbulence becomes $k_\perp=2k_0\sin{\theta_0}$, where $k_0$ is the vacuum wavenumber of the launched wave. Once the value of $n_{co}$ is known for a given beam frequency, an independent measurement of the density profile is also required in order to infer its radial position. In a more realistic setting, density profiles must be used as an input for a ray tracing code in order to calculate the evolution of the refraction index and the incidence angle of the beam. Along this work, such density profiles were obtained using the Thomson scattering (TS) diagnostic \cite{Pasch16} and the Travis ray-tracing code \cite{Marush14} was used to perform the calculation of $k_\perp$, $n_{co}$ and the cutoff position.\\

\begin{figure}
	\centering
	\includegraphics[width=0.5\linewidth]{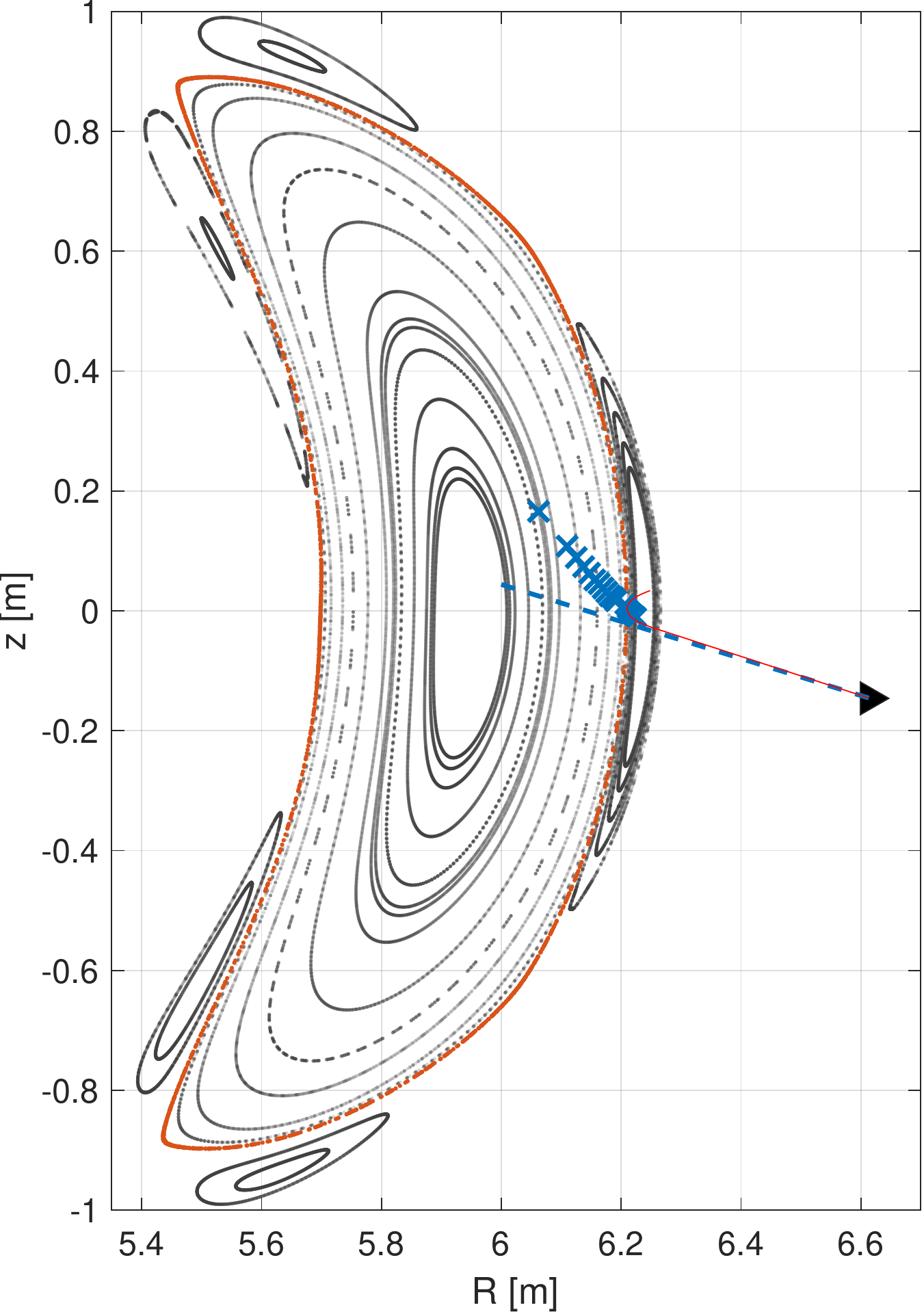}
	\includegraphics[width=0.5\linewidth]{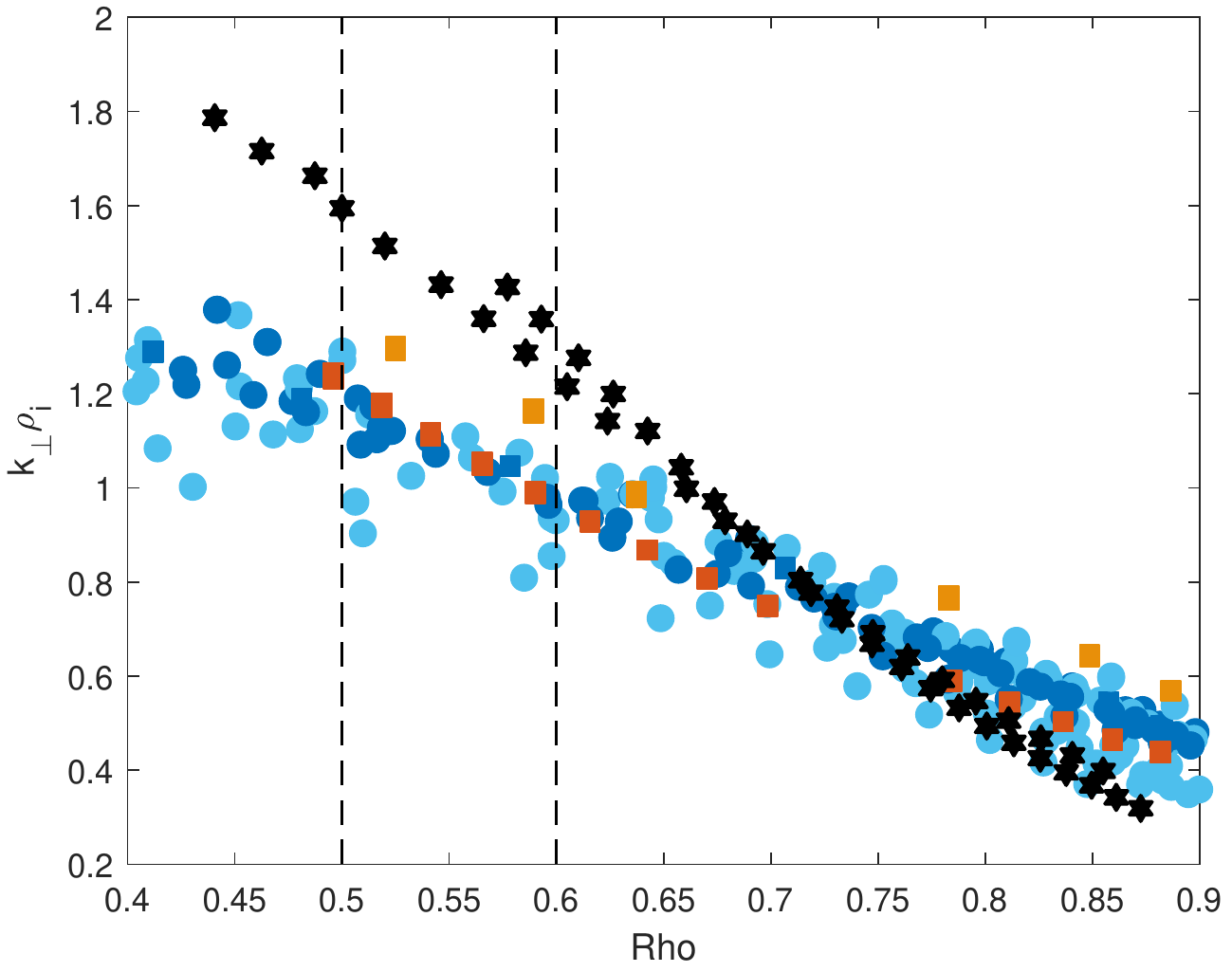}
	\caption{\textit{Top: bean cross section of the standard configuration. The LCFS is indicated as a dotted red line. The DR antenna position is represented as a black triangle and the beam launching direction as a dashed line. Typical measurement points for the $0.4 < \rho < 1.05$ range are indicated as blue crosses. The result of one of the ray tracing calculations is included as a reference as a thin red line. Bottom: $k_\perp\rho_i$ values of the density fluctuations measured by the DR. Data points are representative of the scenarios discussed in section \ref{data}, indicated by colors and symbols as in figure \ref{fig1}. The measurement region $\rho = 0.5-0.6$ is highlighted with dashed lines. }}
	\label{fig00}
\end{figure}

For this work, we will use the DR system installed in the AEA-21 port of W7-X (toroidal angle $\phi = 72^{\circ}$) with the fixed-angle antenna below the equator \cite{Windisch15}. This monostatic DR features a combination of V-band frequencies ($50$-$75$ GHz) and O-mode polarization, which yield cutoff densities of $ 2.8 \cdot 10^{19} $ m$^{-3} < n_{co} < 6.3 \cdot 10^{19}$ m$^{-3}$ in the slab approximation, as indicated by equation (\ref{eq0}). These values mean that access to a wide radial range of positions from the SOL to the core is possible under most experimental scenarios, typically obtaining measurements for $0.5 < \rho < 1$ if core plasma densities are not particularly high or low, where $\rho \equiv (\Psi/\Psi_{LCFS})^{1/2} $ is the radial magnetic coordinate defined using the toroidal flux through a given magnetic surface, $\Psi$, and LCFS stands for last closed flux surface. Under typical operation, the whole frequency range is swept over time, with $10$ ms steps increasing the frequency by $1$ GHz. This means that a full radial profile is produced every $250$ ms. The AEA-21 port is located at the bean-shaped cross section of the plasma and the launching position of this DR leads to measurement points near the outer midplane, where ITG turbulence is expected to be most unstable \cite{Xanthopoulos14,Banon20}. In figure \ref{fig00}, a view of this cross section is provided, along with a typical radial profile of measurement points, the position and launching angle of the antenna and one beam trajectory for reference. For this diagnostic, typical $k_\perp$ values of the backscattering turbulence are in the $k_\perp \simeq 7-11$ cm$^{-1}$. Since the launching angle is fixed, these values depend mostly on the beam frequency, and therefore are also linked to the cutoff-layer position. This is represented as well in figure \ref{fig00}, where the local values of $T_i$ are used to calculate the $k_\perp\rho_i$ values of the measured turbulence for a number of discharges representative of the scenarios which will be discussed in the study. As can be seen, $k_\perp\rho_i \simeq 1-1.5$ is achieved in the highlighted $\rho = 0.5-0.6$ region used for the analysis (as will be explained in the next section). As advanced in the Introduction, this $k_\perp\rho_i$ value represents a good measurement point for ion-scale turbulence and for ITG modes in particular. \\

 Besides the DR, data from other diagnostics were required for this study. First, an interferometer \cite{Brunner18} is used to calculate the plasma density averaged over a line crossing the core, $\bar{n}_e$, in order to provide a reference for the global density of the plasma. Then, the TS system is used to obtain $n_e$ and $T_e$ profiles. As already explained, density profiles are required for the radial localization of DR measurements, and both kinds of profiles are used to obtain local plasma parameters and normalized gradients, $a/L_n$ and $a/L_{T_e}$. In order to obtain $T_i$ profiles, from which to derive $a/L_{T_i}$, two different diagnostics are used: first, the X-ray imaging crystal spectrometer (XICS) is used routinely to obtain $T_i$ measurements from the plasma core to most of the edge by fitting argon impurity spectral lines \cite{Bitter10}. As well, when NBI power is injected (either as means of plasma heating, or in short blips for diagnostic purposes), the charge-exchange recombination spectroscopy system (CXRS) becomes available \cite{Ford20} and provides $T_i$ and rotation data (only the former is used in this study). The agreement between the two kinds of $T_i$ profiles and with $T_e$ data is generally good: electron and ion temperatures have similar values and gradients at the edge. Moving towards the core, both profiles depart at a a certain radius, $\rho_T$ (typically in the $0.5 < \rho_T < 0.75$ range, depending on plasma parameters) after which the $T_i$ profile becomes flatter. When combining all diagnostics to provide a coherent picture, some adjustments have been found to be required. First, in ECRH discharges all heating power is deposited in the electron species and then transferred to the ions via collisions. For this reason, profiles for which $T_i > T_e$ are considered unphysical (not so in NBI discharges, for which that situation is indeed feasible). Therefore, in this case $T_i$ values are substituted by $T_e$ ones whenever this situation is found. This leads sometimes to a problem when using XICS profiles, which are found to overestimate $T_i$ by around $100-200$ eV over the whole profile. Because of this, the two profiles do not join smoothly, leading to discontinuities in the $a/L_{T_i}$ values. However, it has been found generally that if the whole XICS profile is shifted to overcome that bias, not only the agreement with the TS data is good for $\rho > \rho_T$, but also the agreement between XICS and CXRS improves notably. Therefore, XICS data is corrected to reach a smooth coupling with the TS $T_e$ profile.\\

\section{Analyzed scenarios}\label{data}

The objective of this study is to provide a systematic characterization of core density fluctuations and to discuss how they are related to the local drive of turbulence. Therefore, as advanced in the Introduction, all selected discharges will have a common configuration  in order to avoid mixing up possible effects associated to the characteristics of the magnetic configuration with those related to the gradient values. The standard configuration \cite{Beidler11} has been selected for this, as it was the most widely used during the campaign, thus providing a substantially larger database than the rest. Within this subset of the experimental results, selected data should ideally be representative of the most relevant operational regimes achieved during the 2018 experimental campaign and cover the widest possible range of plasma parameters. However, as discussed in the previous section, the DR imposes certain limitations on that range, as it can only measure fluctuation amplitude values within a range of operational parameters for which the core region is accessible for a  given frequency band and polarization. This parametric space is presented in figure \ref{fig1}, in which W7-X operational limits are indicated as two solid lines: the first represents the maximum ECRH power typically available during the campaign, P$_{ECRH,max} \simeq 7$ MW \cite{Klinger19}. This limit is exceeded by a few discharges, as up to $3.5$ MW of NBI could be injected on top of the P$_{ECRH}$. The second is the critical density limit, which has been calculated taking from \cite{Fuchert20} the boronized wall values (featuring a low impurity fraction, $f_{imp} = 0.5\%$) and considering as a profile shape factor\footnote{The profile shape factor, $\delta$, is a parameter introduced in \cite{Fuchert20} which relates the edge density and the line-averaged density, as measured by the interferometer.} the average value for ECRH discharges, $\delta \simeq 1.33$. As well, the density limits to DR operation are indicated by shadowing both regions below the lowest cutoff density (which can not be probed by the V-band) and those for which the highest frequencies can only probe the edge of the plasma. Finally, in order to carry out a meaningful comparison of fluctuation levels between different plasma regimes, a common and relatively narrow radial region must be selected, thus imposing further constrains in the shape of the density profile. Taking into account the availability of data and the localization of the observed changes in fluctuation profiles (which will be discussed in the next section), the region $ 0.5 < \rho < 0.6$ is selected, and only discharges including data in this range are considered in the study. Once all these constrains are taken into account, a database of more than $150$ points from $18$ different discharges is created. As can be seen in figure \ref{fig1}, measurements (indicated as colored symbols) are selected to cover most of the accessible region, specially taking into account that the maximum theoretically available ECRH power was rarely achieved, with most discharges featuring lower power levels in the P$_{ECRH} \leq 6$ MW range.  On top of that, a few extra measurements have been selected as representative of scenarios in which the $T_{i,core}$ clamping is relaxed. This way, the database is divided into three main families of discharges which will be discussed separately along this work:

\begin{figure}
	\centering
	\includegraphics[width=0.65\linewidth]{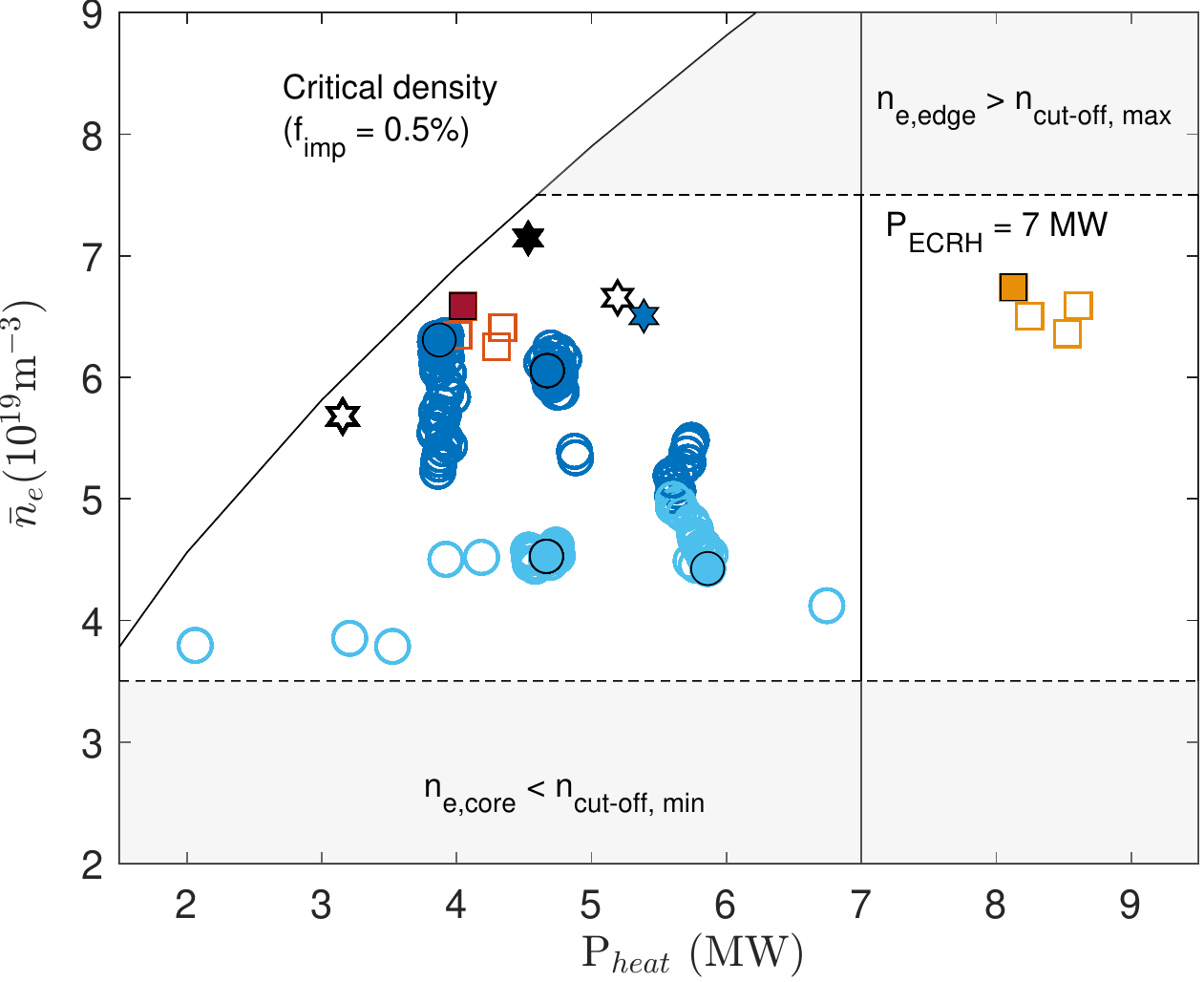}
	\caption{\textit{Parametric map of the experimental data. Measurement points are displayed as a function of the total heating power (including ECRH and NBI) and line averaged density. Four major groups of discharges are identified by symbols and colors: dark/light blue circles represent high/low density gas puff ECRH discharges. Redish squares represent NBI discharges. Black stars represent pellet-fueled high performance discharges. Blue star represents reference ECRH, gas puff phases for the corresponding HP scenario. Solid symbols correspond to the highlighted cases. Solid/dashed lines represent several operational/instrumental limits. }}
	\label{fig1}
\end{figure}

\begin{enumerate}
	\item [-] Gas puff-fueled, ECRH-heated discharges. These represent the bulk of the 2018 experiments and are represented as blue circles. Due to differences in the behaviour of fluctuations which will be introduced in the next section, dark/light colors are used to designate data points above/below a density value of $ \bar{n}_e \simeq 5 \cdot 10^{19}$ m$^{-3}$, which will be in the following referred to as $\bar{n}_0$. These two subgroups will be referred in the following as "high/low density ECRH".
	\item [-] High $T_{i,core}$, NBI-heated discharges. These discharges represent the aforementioned scenario in which $T_{i,core} > 1.7$ keV is achieved by a particular combination of NBI and ECRH heating. They are represented as reddish squares, with an orange/dark red color indicating full/reduced ECRH power in addition to NBI heating (see figure \ref{fig02} below). 
	\item [-] Pellet-fueled, ECRH heated, high performance discharges. These data points represent the transient high confinement regime found after pellet fueling, also reaching $T_{i,core} > 1.7$ keV and already discussed in the Introduction. They are represented as black stars. Blue stars correspond to conventional ECRH discharges with similar $\bar{n}_e$ and $P_{ECRH}$ values, used as a reference.
\end{enumerate}

In order to facilitate the description of these families of discharges, a few representative cases have been selected for each (indicated in figure \ref{fig1} as solid symbols). These examples will be used to guide the discussion and to provide more details when a full representation of the whole database would be impractical. Next, these particular discharges are presented and used to show paradigmatic data from each of the groups.

\subsection{ECRH discharges}

\begin{figure}
	\centering	
	\includegraphics[width=0.45\linewidth]{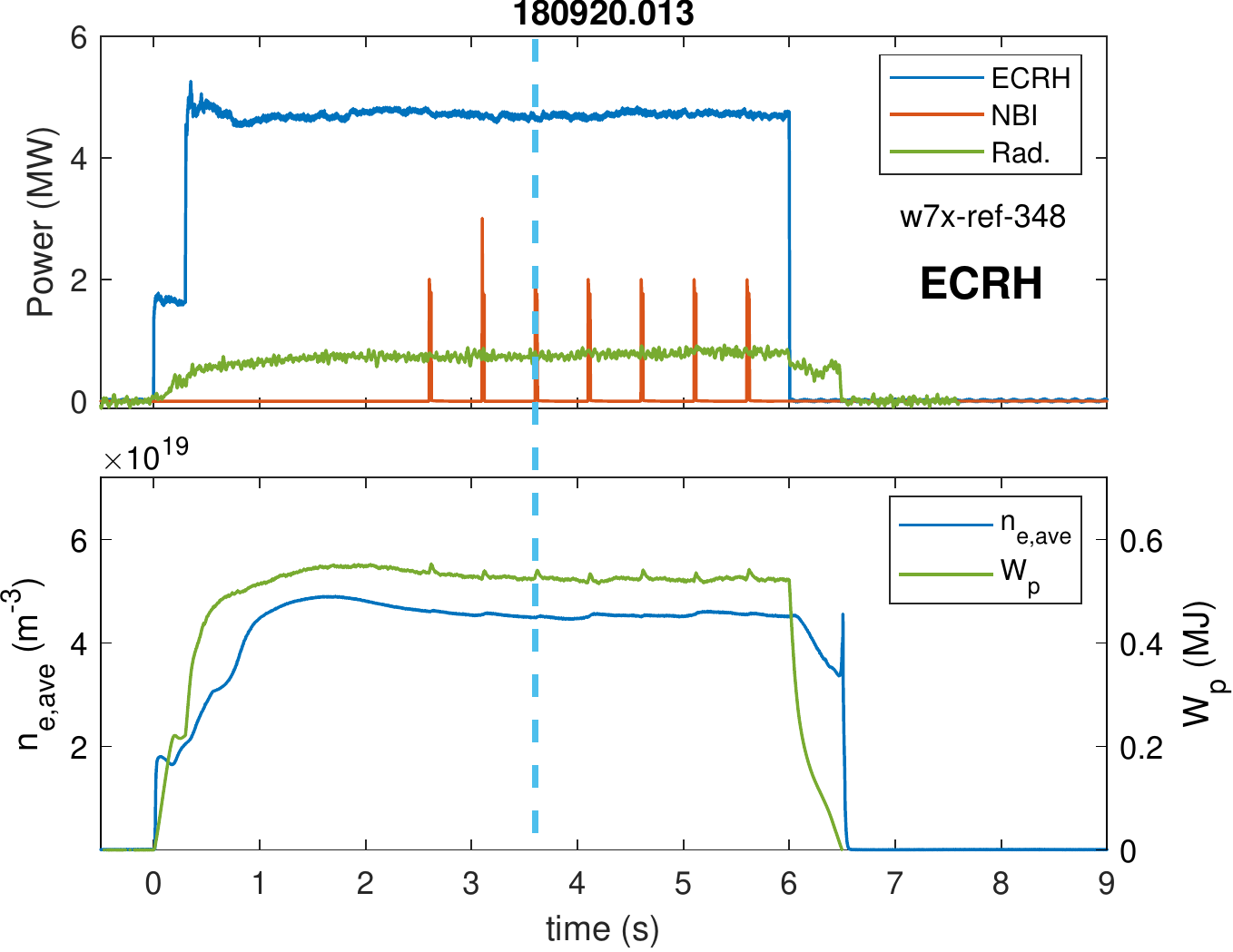}
	\includegraphics[width=0.45\linewidth]{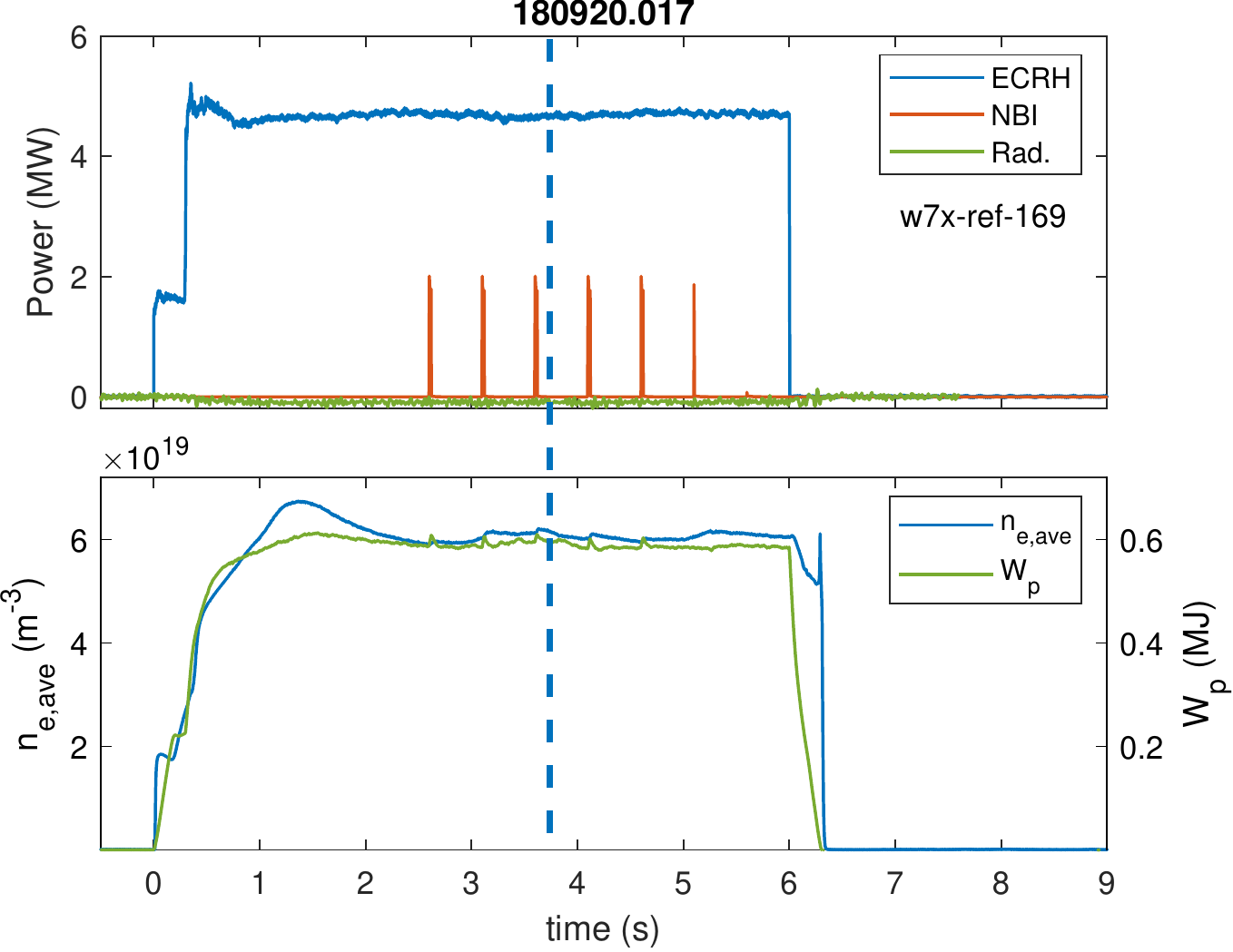}
	\par
	\includegraphics[width=0.45\linewidth]{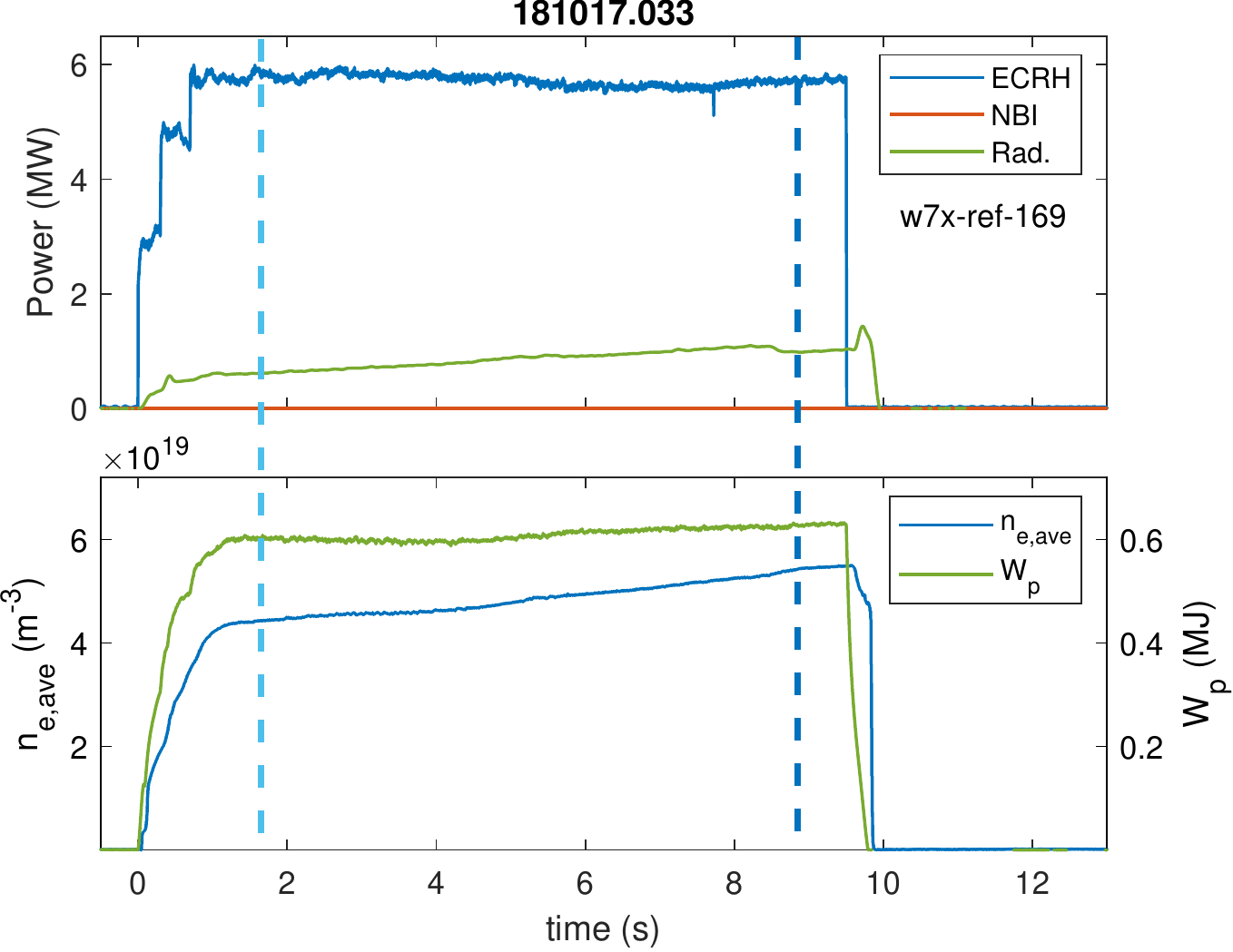}
	\includegraphics[width=0.45\linewidth]{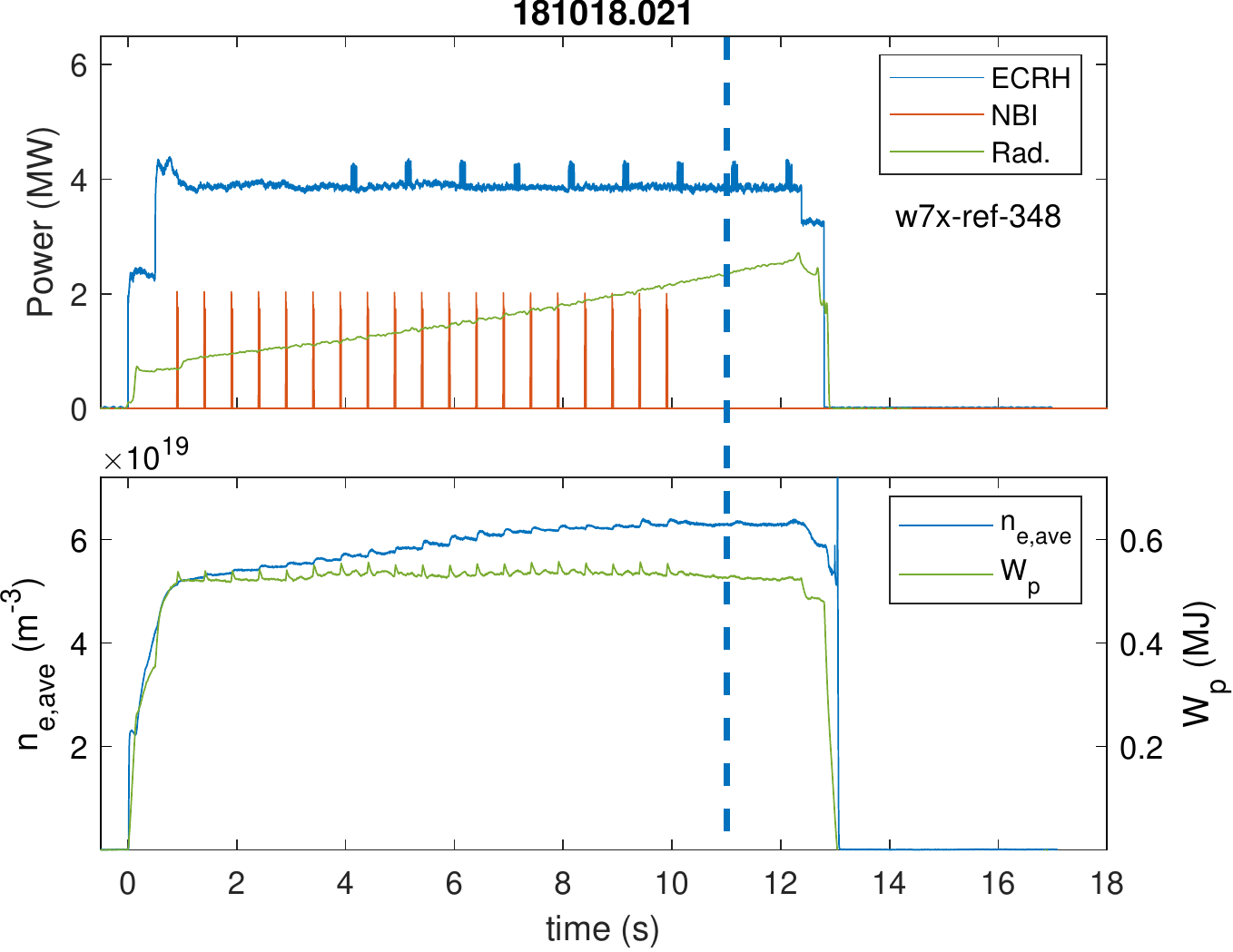}
	\caption{\textit{Representative cases of gas-puff, ECRH discharges. On the top plot of each graph, temporal evolution of heating power and radiated power are displayed. On the bottom plot, line averaged density (blue, left axis) and stored energy (green, right axis) are shown. Dashed lines indicate the times corresponding to the symbols of the same colors in figure \ref{fig1} and successive.}}
	\label{fig01}
\end{figure}

Four discharges have been selected to represent the first group, all of which are shown in figure \ref{fig01}. The first two are $180920.013$ and $180920.017$, use the same heating power ($P_{ECRH} \simeq 4.7$ MW) and feature constant density values respectively below and above $\bar{n}_0$. As well, short NBI blips are injected for instrumental purposes, thus allowing for the use of CXRS in the corresponding times. Theses discharges are represented in the plots at the top row of the figure. Instead, $181017.033$ and $181018.021$ are density ramps, in which $\bar{n}_e$ is increased for the duration of the discharge. Discharge $181017.033$, displayed in the bottom left of the figure, features high heating power ($P_{ECRH} \simeq 6$ MW) and crosses the $\bar{n}_0$ limit around one half into the shot. One time is selected for each part of the discharge in this case. Instead, $181018021$ (bottom right) features lower heating power ($P_{ECRH} \simeq 4$ MW) and $\bar{n}_e > \bar{n}_0$ at all times. In these last two shot, the representative point has been selected at time corresponding respectively to the highest density value. Given the interest of these shots, a large number of frequency ramps have been analyzed for each of them (10-15 for each discharge), leading to a corresponding cloud of points around the solid symbol in figure \ref{fig1}. In the case of the density ramps, these clouds extend vertically, as they cover the full range of density values found in the discharge.

\subsection{NBI discharges}

\begin{figure}
	\centering	
	\includegraphics[width=0.45\linewidth]{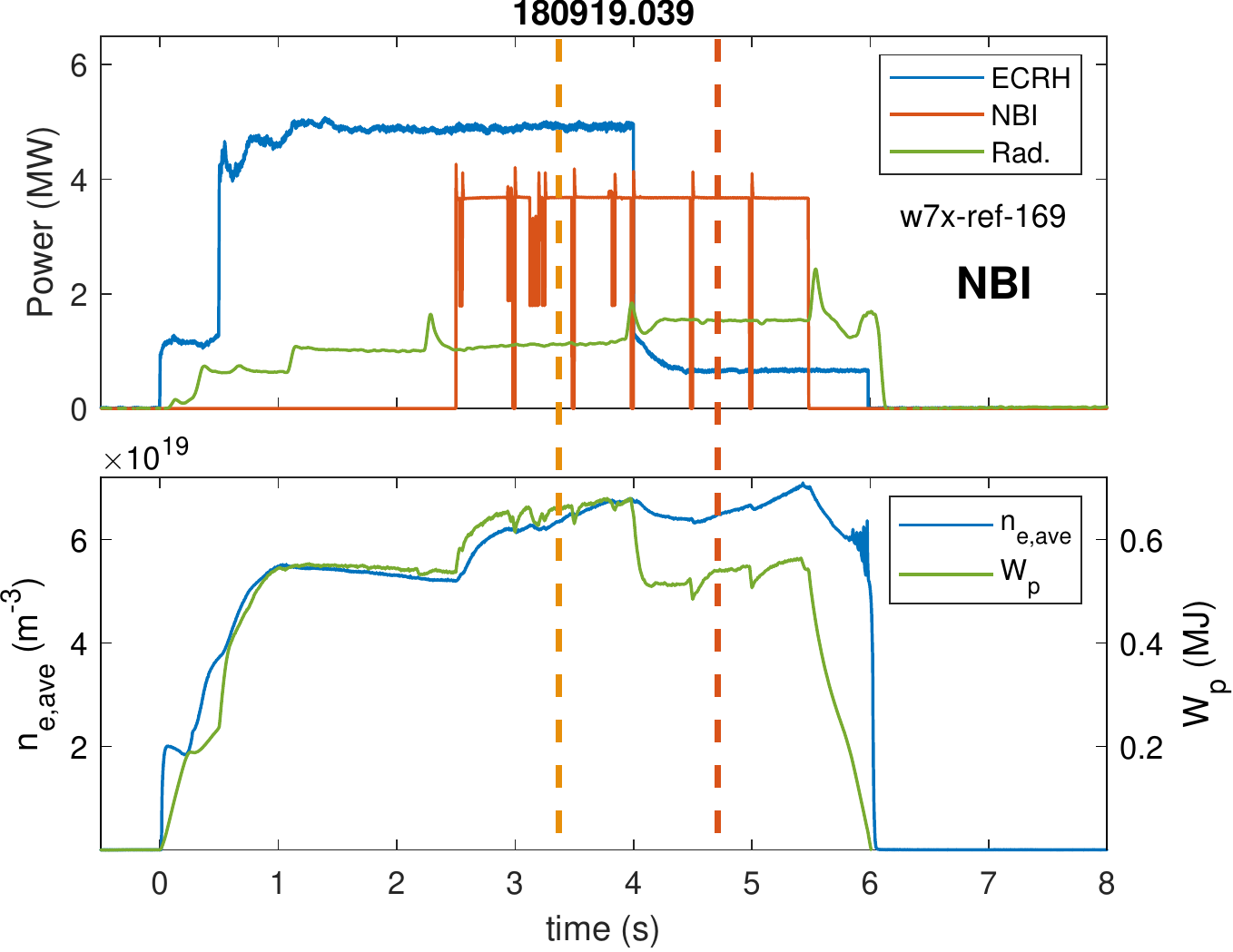}
	\caption{\textit{Representative NBI heated discharge. Plots and colors as in figure \ref{fig01}. Orange and red dashed lines indicate the times corresponding to the symbols of the same colors in figure \ref{fig1} and successive.}}
	\label{fig02}
\end{figure}

The two NBI heating regimes shown in figure \ref{fig1} correspond to different times of a family of discharges in which different levels of ECRH and NBI are alternated with a relatively stable density. Discharge $180919.039$ is presented in figure \ref{fig02} as an example of this, with dashed lines indicating the two times corresponding to the symbols of the same color. As can be seen, first there is a phase with only ECRH heating, featuring parameters not too far from those of $180920.017$ (top right plot of figure \ref{fig01}), although with slightly lower density. Then, for $2.5$ s $ < t < 4$ s, the full $3.5$ MW of available NBI power are injected into the plasma for a total nominal heating power in excess of $8$ MW (in fact, nominal NBI values may somewhat overestimate the actual power being deposited in the plasma \cite{Lazerson20}), causing a moderate increase of density and stored energy. Finally, for $t > 4$ s there is a third phase in which the NBI heating is conserved, while ECRH heating is reduced to only $P_{ECRH} \simeq 0.5$ MW. In this phase, density gradient is increased across the whole profile and $T_{i,core}$ slightly over the clamping limit is achieved. Nevertheless, as density rises later in the discharge ($t > 5$ s) this transient phase ends and $T_{i,core}$ drops again under the limit.

\subsection{High performance discharges}

\begin{figure}
	\centering		
	\includegraphics[width=0.45\linewidth]{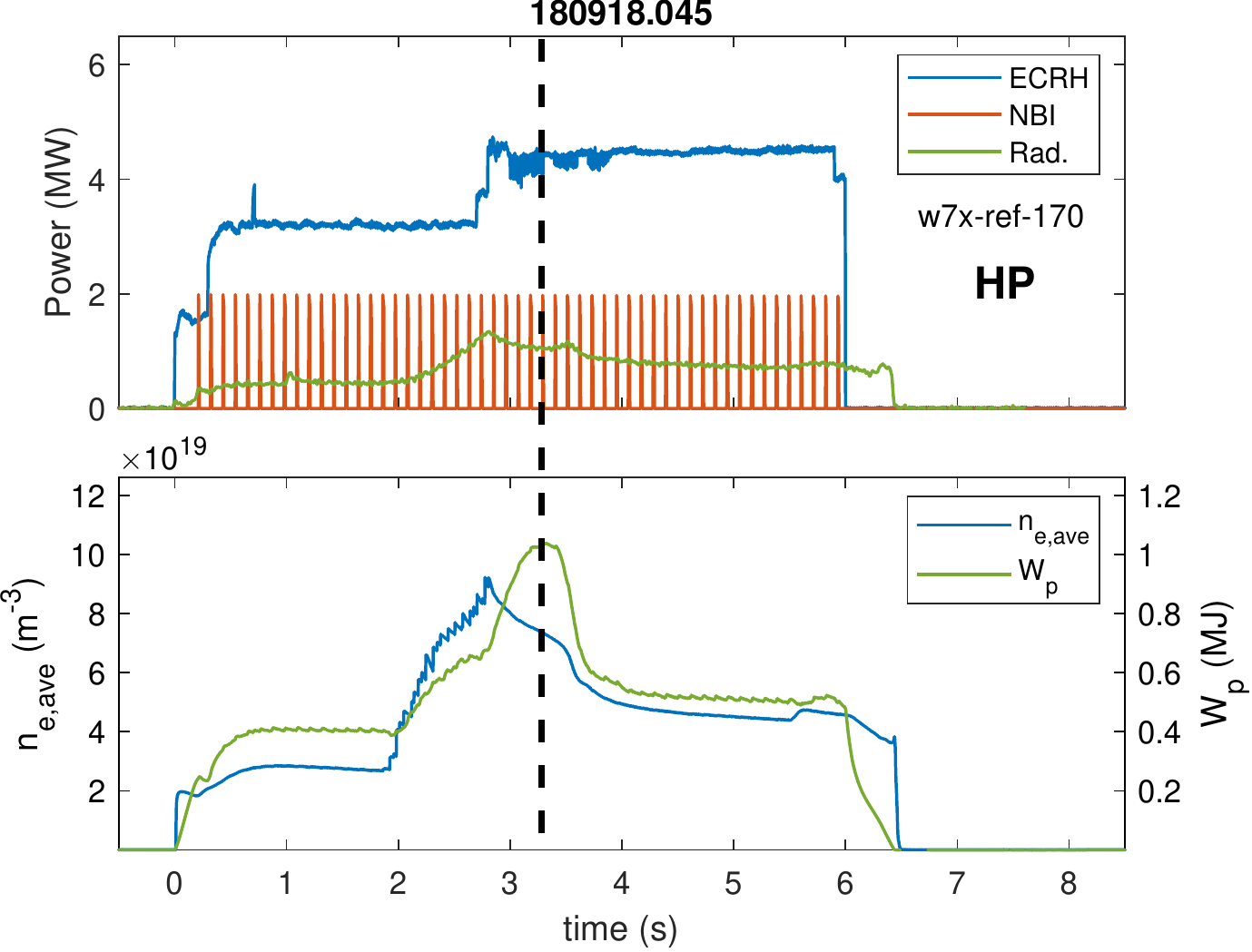}	
	\includegraphics[width=0.45\linewidth]{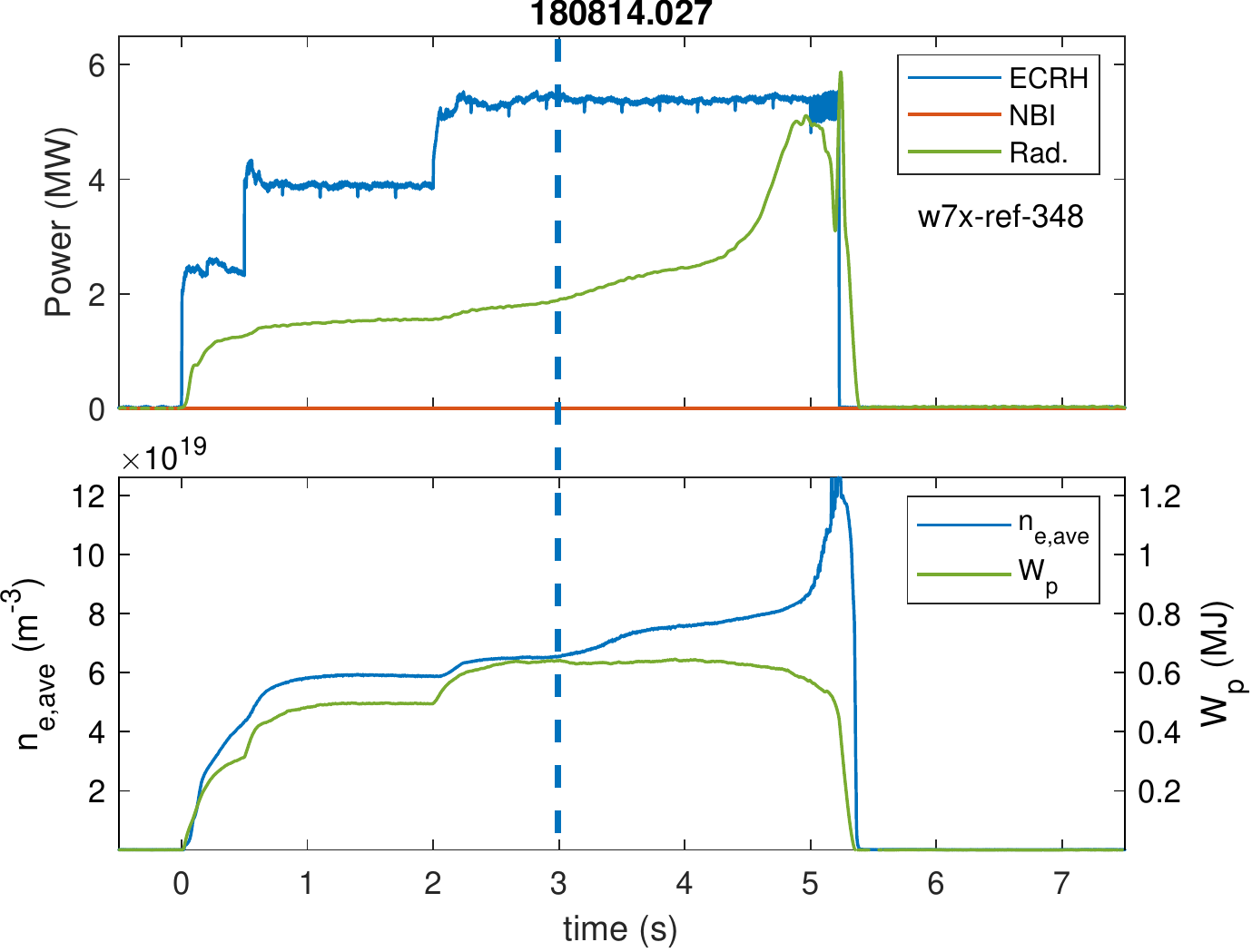}
	\caption{\textit{Representative high performance discharge (top graphic) and reference ECRH shot (bottom). Plots and colors as in figure \ref{fig01} (mind the different scale in the y-axis in the lower plots). Black and blue dashed lines indicate the times corresponding to the symbols of the same colors in figure \ref{fig1} and successive.}}
	\label{fig03}
\end{figure}

High performance (HP) discharges have already been described at length in the literature \cite{Bozhenkov20}, including a dedicated study of the DR data which already showed how this regime is characterized by a strong suppression of turbulence and a steep E$_r$ well at the plasma edge \cite{Estrada21}. In this work, we take a few representative examples from that study, in which several discharges featuring a HP phase were compared to conventional gas-puff ECRH ones with equivalent density and $P_{ECRH}$ values. One of such comparisons is presented in figure \ref{fig03}: on the top plots, $180918.045$ features the characteristic pellet-induced, sharp increase of density after $t = 2$ s, followed by the HP phase: a fast, transient improvement of $W_p$ and $\tau_E$ after the end of the pellet injection. In the bottom plots, a gas-puff fueled discharge with a density ramp equivalent to the high density ECRH shots previously discussed. In it, a time is selected such that values close to those of the HP phase in $180918.045$ are achieved in both $\bar{n}_e$ and $P_{ECRH}$.


\section{Experimental results}\label{exp}

\begin{figure}
	\centering	
	\includegraphics[width=0.32\linewidth]{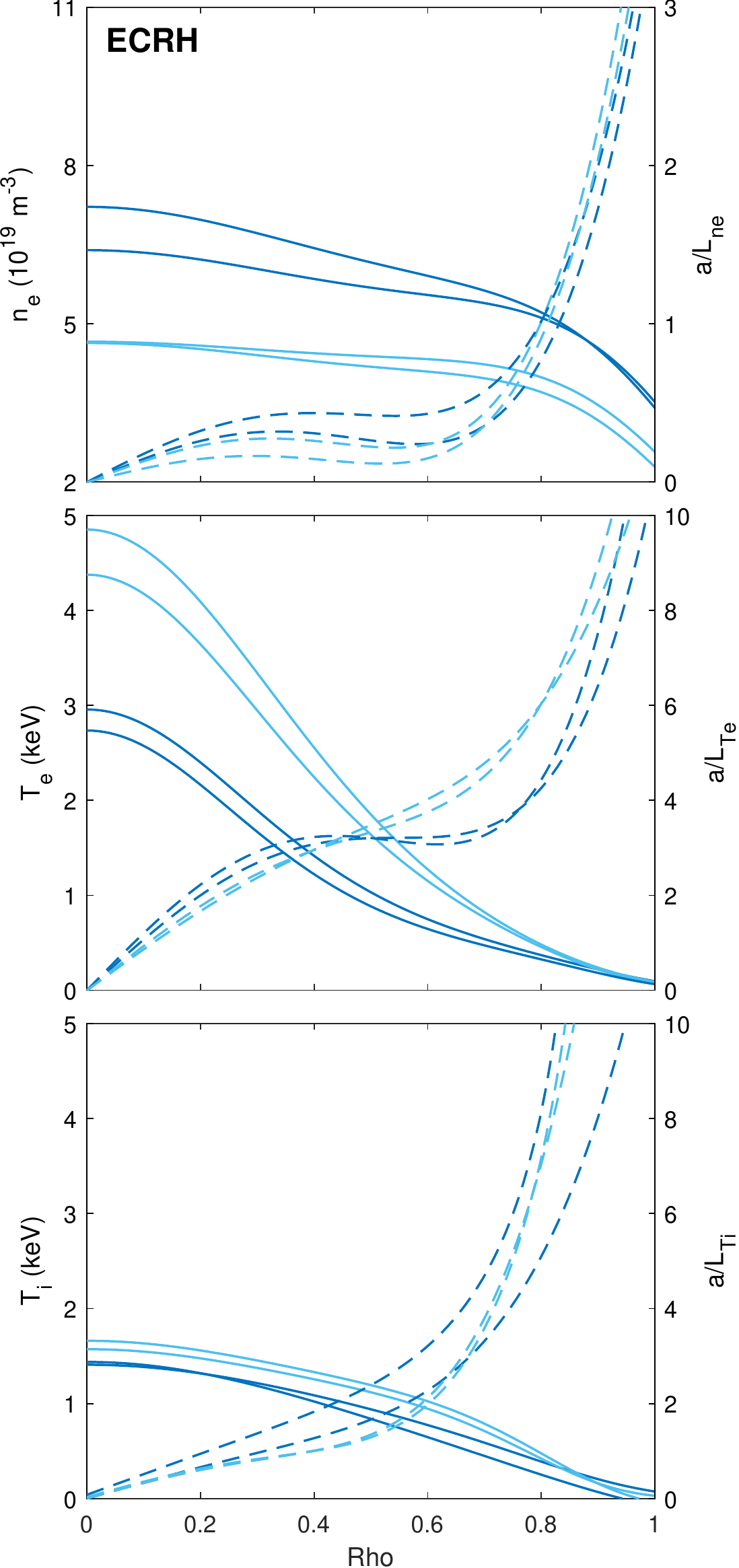}
	\includegraphics[width=0.32\linewidth]{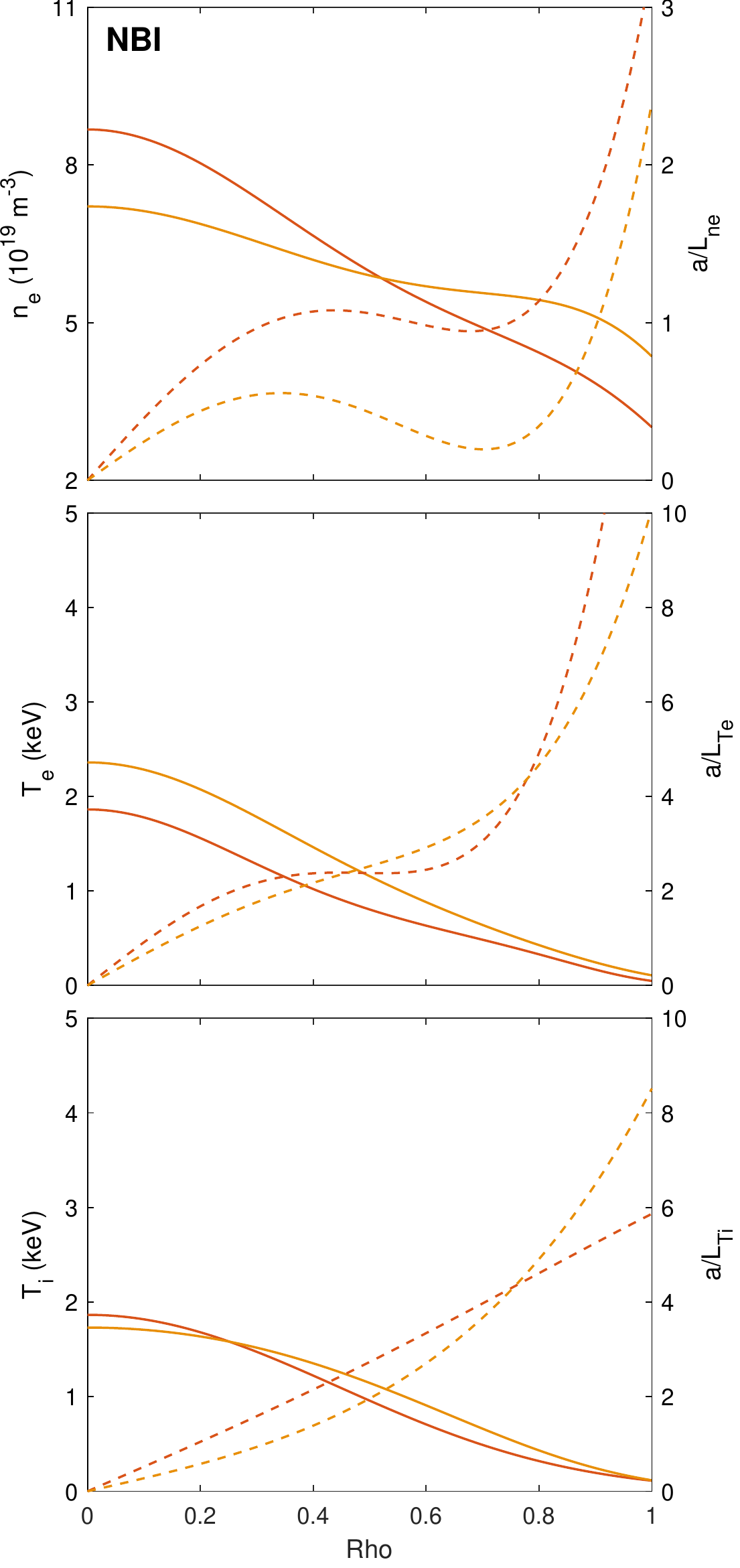}
	\includegraphics[width=0.32\linewidth]{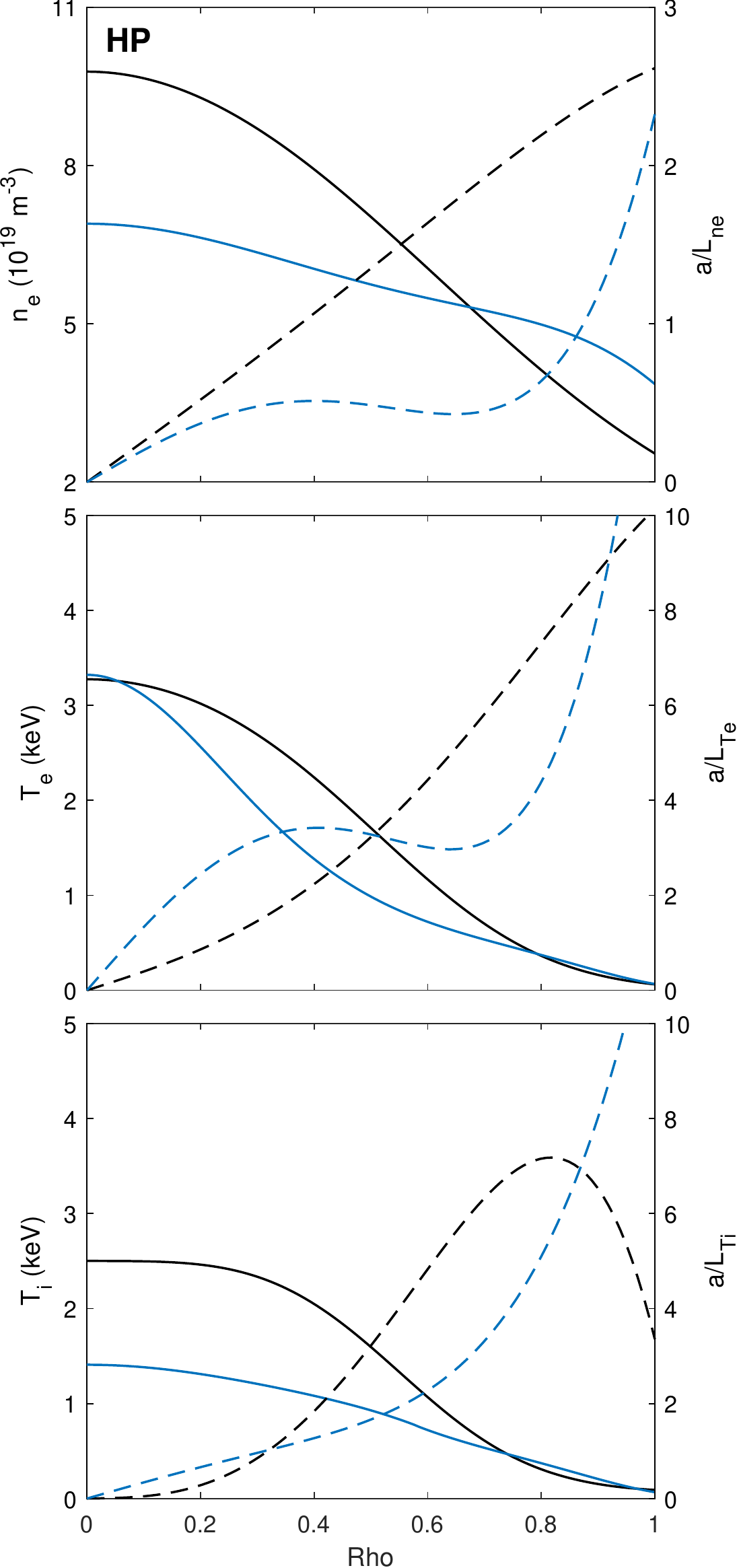}
	\caption{\textit{Typical density and temperature profiles of the representative discharges. From top to bottom, each row represents the profiles (solid) and normalized gradients (dashed) of $n_e$, $T_e$ and $T_i$. Each column corresponds to a different family of discharges including, from left to right, ECRH, NBI and HP. The HP family includes its reference ECRH phase in blue. Colors correspond to the symbols in figure  \ref{fig1}.}}
	\label{fig2}
\end{figure}

The families defined in the previous section will be used as reference to describe the different experimental regimes observed in the database, often referring to the representative discharges highlighted in figure \ref{fig1}. First, typical density and temperature profiles are presented in figure \ref{fig2}: in it, data from each family are displayed in a separate column, including density and electron/ion temperature profiles and normalized gradients, $a/L_\alpha$, $a$ being the plasma minor radius, measured by TS, CXRS and XICS, as discussed in section \ref{Met}. Each curve stands for one of the representative discharges, sharing color with the corresponding symbol in figure \ref{fig1}. As can be seen, density profiles are typically flat towards the core, featuring $a/L_{n_e} \leq 0.5$ for $\rho < 0.7$. Instead, a different situation can be seen in the third phase of the NBI discharges (in red) and during the HP phases, in which the density profile becomes substantially more peaked. As well, it can be seen how as density increases in ECRH scenarios, the shape of the profile (and thus the gradients) becomes slightly more peaked, but does not change substantially. Electron temperature is the parameter with a greatest degree of external control, as it depends directly on $P_{ECRH}$ and density, reaching core values of several keV and up to $T_{e,core} \simeq 5$ keV for the low density ECRH shots in figure \ref{fig2}. As a result, $T_e$ gradient reaches rather high values in all cases, with  $a/L_{T_e} \simeq 3-4$  for $\rho = 0.5-0.6$. Finally, $T_i$ profiles contrast with those of $T_e$ as a result of the clamping already discussed in the Introduction: with a few exceptions, $T_i$ profiles are similar to those of $T_e$ at the edge, but then become flatter at the core ($a/L_{T_i} \leq 2$ for $\rho = 0.5-0.6$) and remain below the core limit value of $T_{i,core} \simeq 1.7$ keV. Again, figure \ref{fig2} show two exceptions to this rule, as already discussed and reported elsewhere in the literature \cite{Bozhenkov20,Ford19}: the third phase of the NBI discharges featuring reduced values of $P_{ECRH}$, in which $T_{i,core}$ reaches 1.9 keV, and the HP phase, in which $T_{i,core}$ values clearly exceeding the clamping value are found.\\

\begin{figure}
	\centering	
	\includegraphics[width=\linewidth]{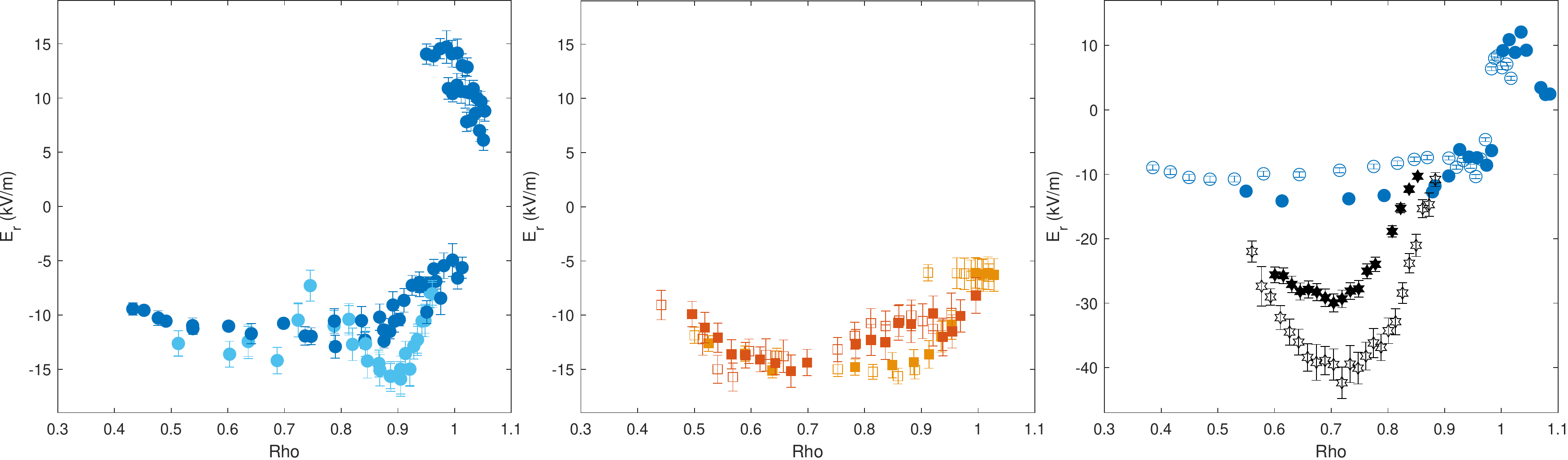}
	\caption{\textit{Radial electric field measurements from the DR. Local $E_r$ values are presented, with no $|\nabla r|$ normalization to account for flux expansion in the magnetic surface \cite{Carralero20}. As in figure \ref{fig2}, each column corresponds to a different family of discharges (from left to right, ECRH, NBI and HP), with colors following the same convention. Solid symbols correspond the representative points of the same colors in figure \ref{fig1}.}}
	\label{fig3a}
\end{figure}

Once the plasma profiles have been presented for the different cases, measurements from the DR can be discussed: in figure \ref{fig3a}, characteristic $E_r$ profiles from each family are displayed. Again, solid symbols correspond the representative points of the same colors in figure \ref{fig1}. As can be seen, $E_r$ profiles in ECRH discharges are consistent with those already presented in previous work, in which good qualitative agreement with neoclassical predictions was found \cite{Carralero20,Estrada21}: negative $E_r$ inside the confined region, in good agreement with the dominant ion-root ambipolar condition at the plasma edge \cite{Hirsch08}, and positive in the SOL. Lower density discharges feature stronger electric field at the edge, with more negative minimum values around $\rho \simeq 0.9$. 
NBI shots show $E_r$ profiles which are remarkably similar to those of high density ECRH discharges in the core. Instead, it can be seen that no field reversal is observed around the separatrix, suggesting a different behaviour at the SOL, with substantially reduced or even negative values of $E_r$. The reason for this is unclear and should be addressed in a separate work. Finally, as already discussed in \cite{Estrada21}, HP phases display substantially stronger negative electric field (note the different scale in figure \ref{fig3a}), which reach their minimum around $\rho \simeq 0.7$. As shown in previous work this value depends on the ECRH heating and density: lowest $E_r$ values can be reached by increasing $P_{ECRH} \cdot \bar{n}_e$ as shown by hollow points corresponding to discharge $180918.041$ (and the corresponding reference in blue), with a $P_{ECRH} \cdot \bar{n}_e$ value roughly $10\%$ greater than that of $180918.045$, represented by solid symbols).\\

\begin{figure}
	\centering	
	\includegraphics[width=\linewidth]{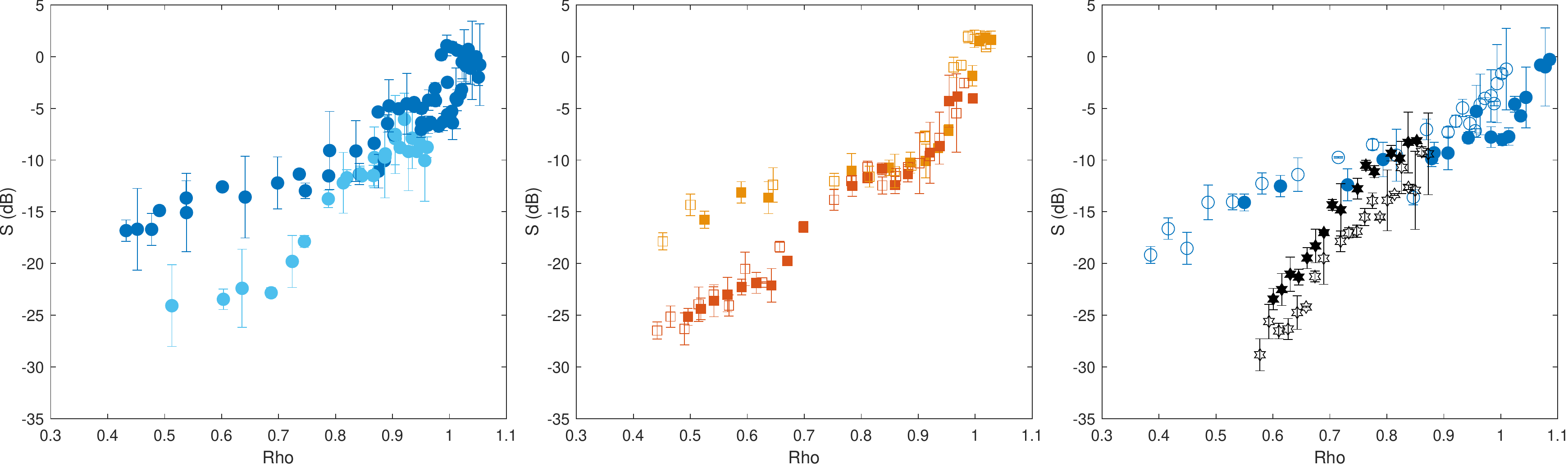}
	\caption{\textit{Fluctuation measurements from the DR. Signal power, $S$ is represented in dB. As in figure \ref{fig2}, each column corresponds to a different family of discharges (from left to right, ECRH, NBI and HP), with colors following the same convention. Solid symbols correspond the representative points of the same colors in figure \ref{fig1}.}}
	\label{fig3b}
\end{figure}

Figure \ref{fig3b} displays the measurements of density fluctuations from the DR following the same convention of symbols, columns and colors as figure \ref{fig3a}. In them, signal power $S \propto \delta n^2$ is represented in dB. As can be seen, fluctuation profiles are similar in all high density ECRH discharges (including the reference phase of the HP family): a sharp decrease in the vicinity of the separatrix followed by a shallow decreasing slope towards the core, reaching values around $S \simeq -15$ dB for the analyzed radial region, $\rho \simeq 0.5-0.6$. It must be taken into account that, given the logarithmic scale of the y-axis, this ``shallow'' slope still implies a drop of an order of magnitude in signal power with respect to the levels at the separatrix. Starting from this baseline case, a reduction in core turbulence amplitude can be seen in three scenarios: first, in the left plot, low density ECRH discharges have similar values at $\rho \simeq 0.9$, but then feature a marked drop of $S$ starting around $\rho \simeq 0.8$ and falling ca. $10$ dB below the previous case at $\rho = 0.5$. A very similar observation can be made in the third phase of the NBI discharges (corresponding to reduced ECRH heating) in the central plot, although interestingly, the intermediate phase -with full ECRH and NBI heating- remains mostly unaffected. Finally, on the right plot, HP discharges display a somewhat sharper suppression of turbulence than the other two scenarios, which may in some cases start at more exterior radial positions $\rho > 0.8$. In this case, signal power in the range of $S \simeq -30$ are achieved. Besides the behaviour at the core, it is interesting to point out that, between the $\rho \simeq 0.9$ region -in which similar $S$ values are found in all scenarios- and the separatrix, different levels of turbulence can be found:  ECRH cases with lower density feature a stronger suppression of turbulence in that region, which suggests a relation to the greater $E_r$ shear present in those cases \cite{Carralero20}. However, a detailed discussion of edge turbulence is out of the scope of this work, and will be addressed in future studies. As well, it must be pointed out that, as indicated in section \ref{Met}, this interpretation of the data involves the hypothesis that turbulence levels are low enough for a linear power response. For higher turbulence levels, the response of the signal to the fluctuations becomes non-linear, increasing substantially the complexity of the analysis \cite{Pinzon17, Happel17}, as the relation between $S$ and $\delta n$ changes non-trivially depending on the region of the $k_\perp$ spectrum being measured, thus invalidating the simple relation in equation (\ref{eq1}). A discussion of the non-linear power response effects on the DR measurements is out of the scope of this work and is left for future studies. In any case, there are two reasons to expect that these effects are not too relevant for the results presented here: first, we will only be discussing measurements carried out in the core region, with radial positions meeting $\rho < 0.6$. In this region, fluctuation levels are expected to be moderate and therefore non-linear effects substantially less likely to become relevant. Second, as can be seen in figure \ref{fig00}, $k_\perp$ values of the discussed fluctuations are very similar due to the narrow radial region being considered. This also reduces the risk of the analysis being affected by non-linear effects, since no comparison is made between different regions of the wavenumber spectrum. \\

\section{Characterization of potential drives for turbulence}\label{drives}

\begin{figure}
	\centering	
	\includegraphics[width=\linewidth]{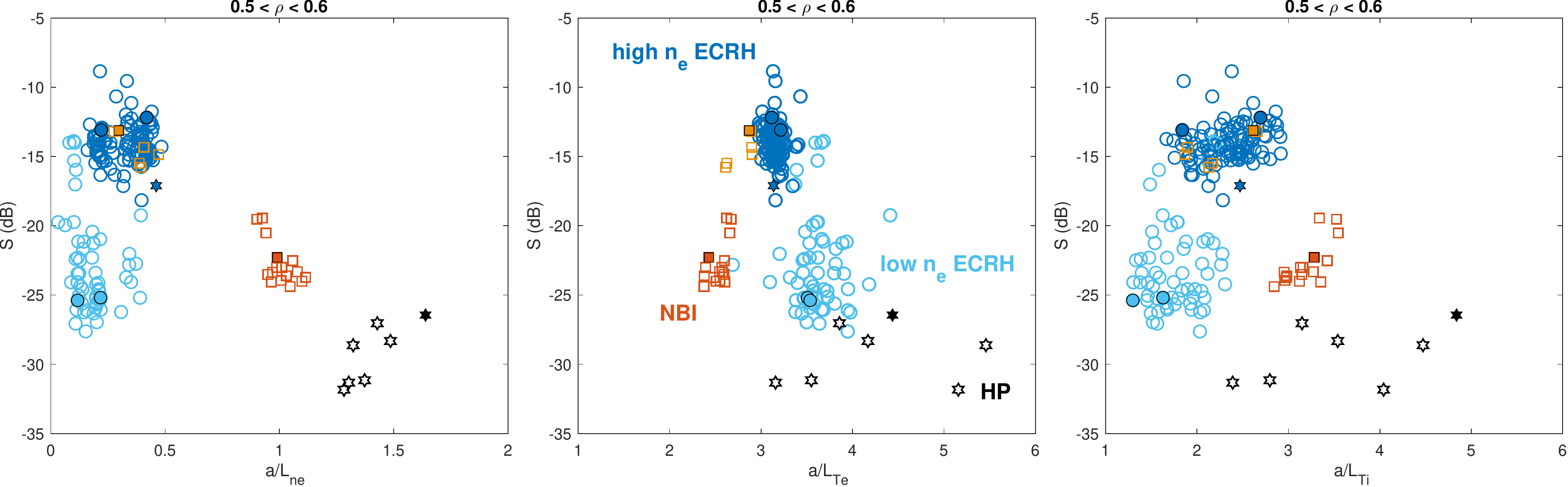}
	\caption{\textit{Core fluctuations as a function of local values of normalized gradients. From left to right, normalized gradients of density, electron temperature and ion temperature. Colors and symbols indicate discharge family and representative shots, as in figure \ref{fig1}. }}
	\label{fig4_a}
\end{figure}

Once fluctuations and profiles have been independently characterized, the relation between them can be addressed by relating fluctuation levels to the different elements potentially driving or suppressing turbulence. In figure \ref{fig4_a} the relation to local gradients is shown, with $S$ measurements from all three families of discharges represented as a function of normalized gradients of $n_e$, $T_e$ and $T_i$. As can be seen, the marked decrease of fluctuations in low density ECRH discharges observed in figure \ref{fig3b} can not be associated to any substantial differences in the gradients with respect to the high density ones: both groups share very similar, low values of density gradient, $a/L_n < 0.5$, with the former featuring higher values of $a/L_{T_e}$ and also slightly lower values of $a/L_{T_i}$. In NBI discharges, the second phase (full NBI+ECRH power, displayed as orange squares) features similar levels of fluctuations to those of high density ECRH discharges, and consistently, present very similar gradient values as that sub-group. Instead, in the third phase (reduced ECRH phase, displayed as red squares), the substantial decrease of fluctuations coincides with a clear increase of $n_e$ and $T_i$ gradients, as well as a more moderate decrease of $a/L_{T_e}$, consistent with the reduction of electron heating. Finally, the strong suppression of turbulence measured in HP discharges follows the trends observed for the NBI discharges in density and ion temperatures, with steeper profiles and higher normalized gradients as well as even lower fluctuation amplitudes.\\

\begin{figure}
	\centering	
	\includegraphics[width=0.5\linewidth]{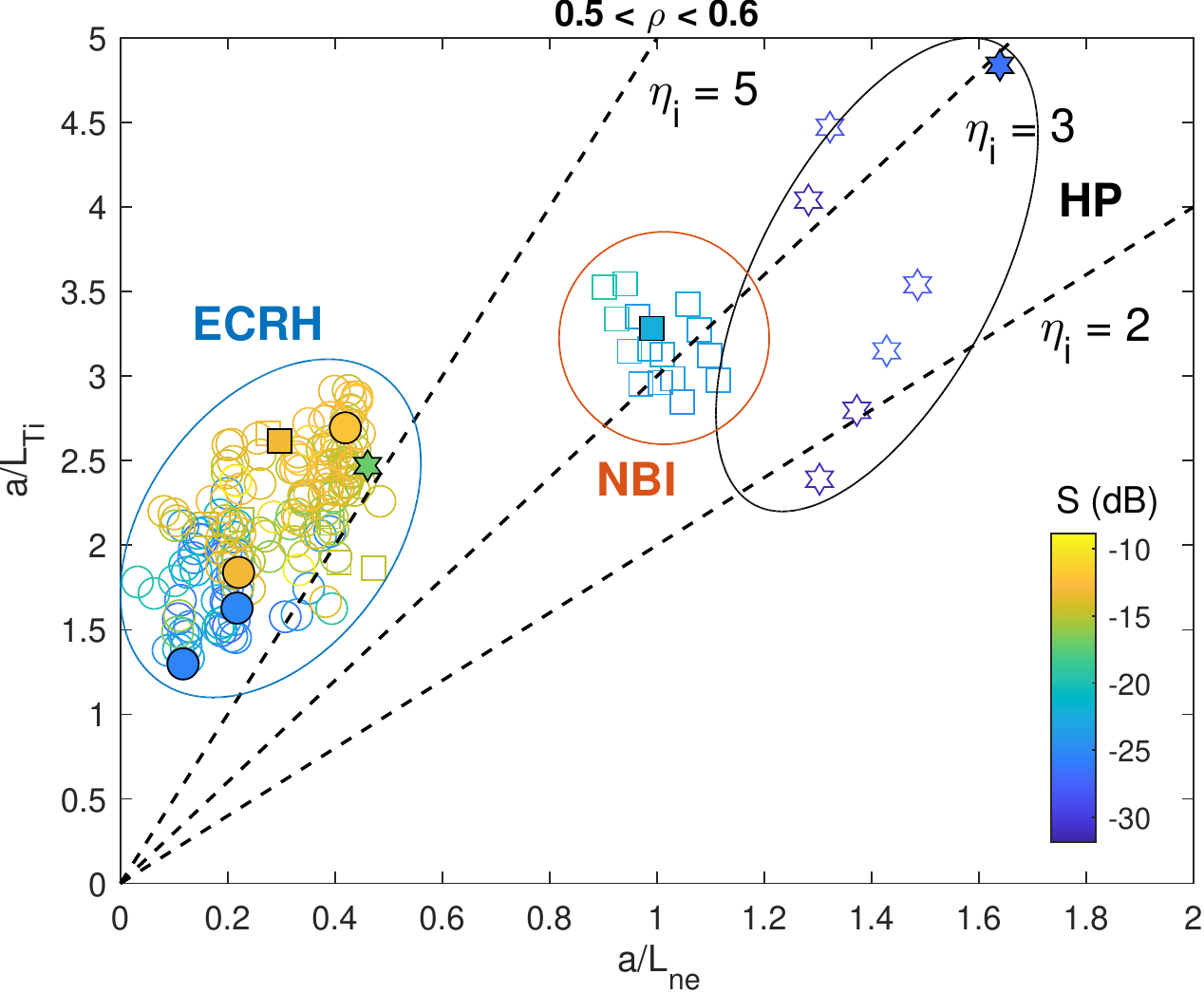}
	\caption{\textit{Fluctuation amplitude (in color), as a function of ion temperature normalized gradients, $a/L_{T_i}$ and $a/L_n$. Symbols represent discharge families (highlighted by colored globes) following previous convention, with solid ones indicating representative discharges. Dashed lines indicate selected values of the $\eta_i = L_n/L_{T_i}$ gradient ratio.}}
	\label{fig4_b}
\end{figure}

These trends become more clear when fluctuations are represented as a function of both $a/L_n$ and $a/L_{T_i}$, as in figure \ref{fig4_b}: as can be seen, each of the families is neatly separated in a different region of the parameter space (highlighted in the figure with colored globes). In general, it can be said that the increase of density gradients for NBI and HP regimes is substantially stronger than that of $a/L_{T_i}$, leading to a clear trend of increasing fluctuations with the gradient ratio parameter $\eta_i = L_n/L_{T_i}$. This fact, highlighted by a few dashed lines indicating selected values of $\eta_i$ in the plot, is perfectly consistent with ITG turbulence expectations, as will be discussed in section \ref{discussion}. There is however, one exception to this general trend, which is represented to the group of low $S$ points in the lower left corner of the ECRH group. As can be deduced from the values presented in figure \ref{fig4_a}, these points correspond to the low density ECRH discharges. In this case, the reduction of fluctuation amplitude seems to be related to a different physical process as their $\eta_i$ values are rather similar, if not greater than the ones corresponding to high density ECRH shots, featuring high $S$ values. This different behaviour will also be addressed in the Discussion.\\

\begin{figure}
	\centering	
	\includegraphics[width=0.5\linewidth]{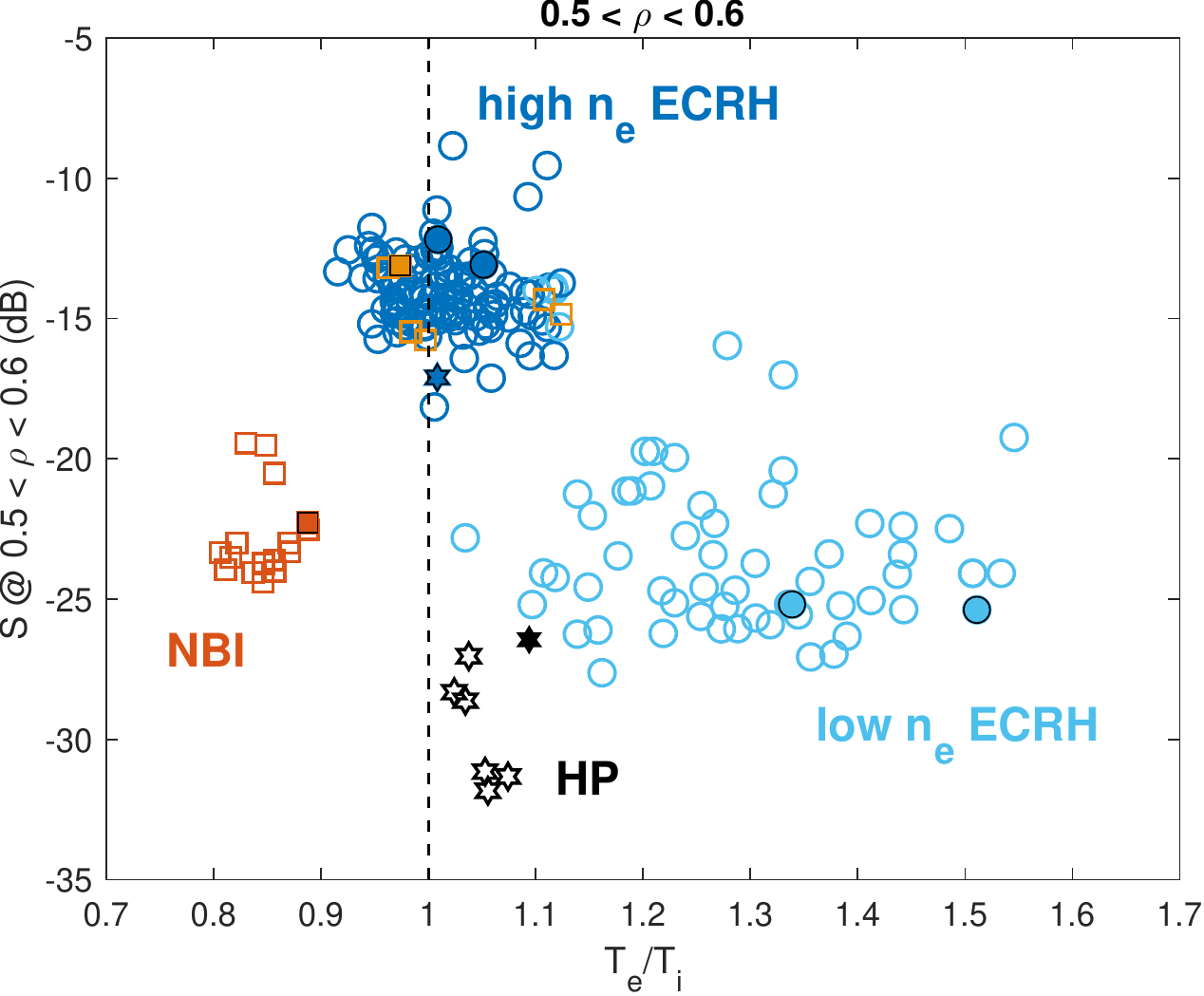}
	\caption{\textit{Fluctuation amplitude as a function of the local electron/ion temperature ratio. Symbols and colors represent discharge families and representative discharges following previous convention.}}
	\label{fig4_c}
\end{figure}

\begin{figure}
	\centering		
	\includegraphics[width=0.5\linewidth]{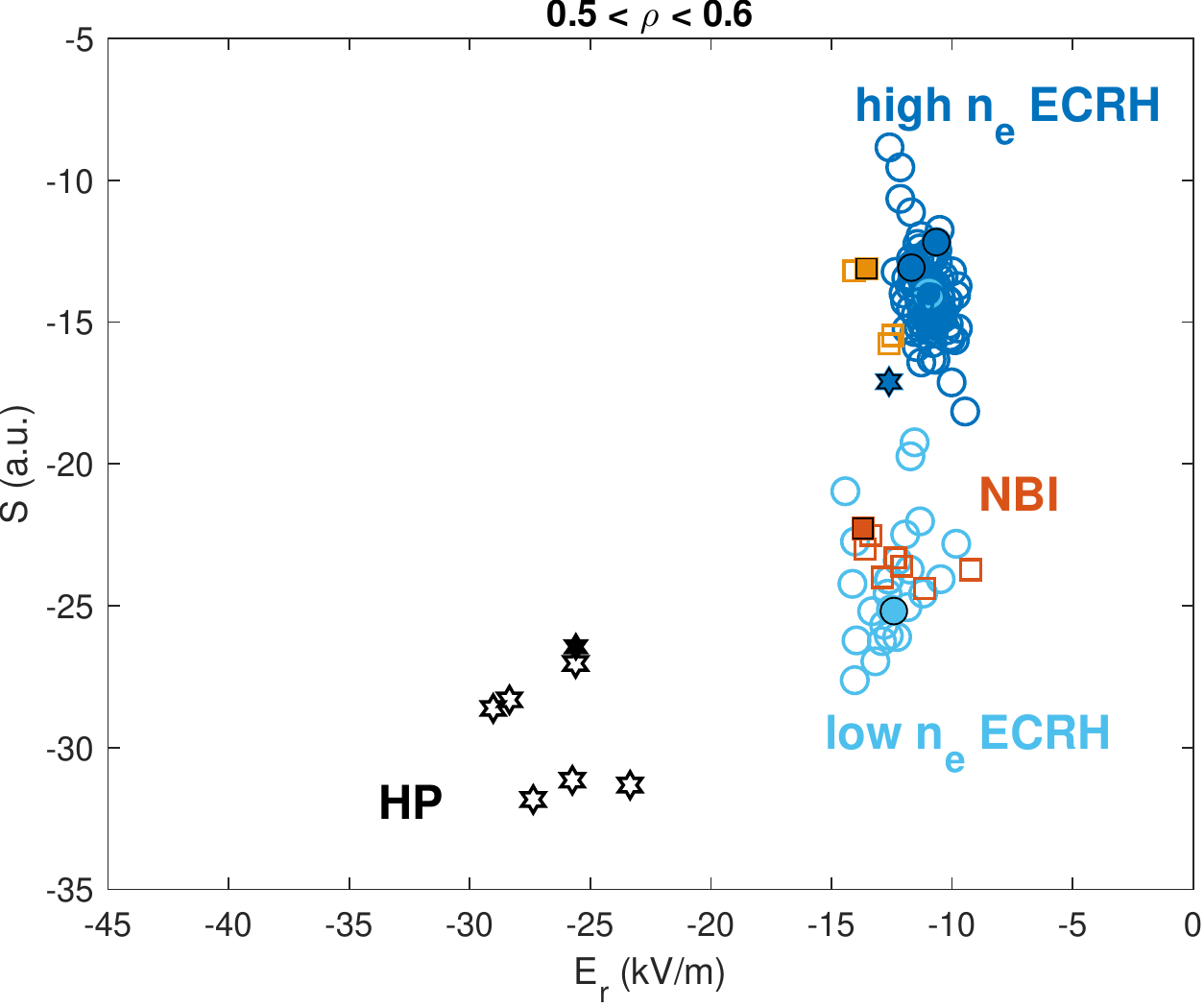}
		\includegraphics[width=0.5\linewidth]{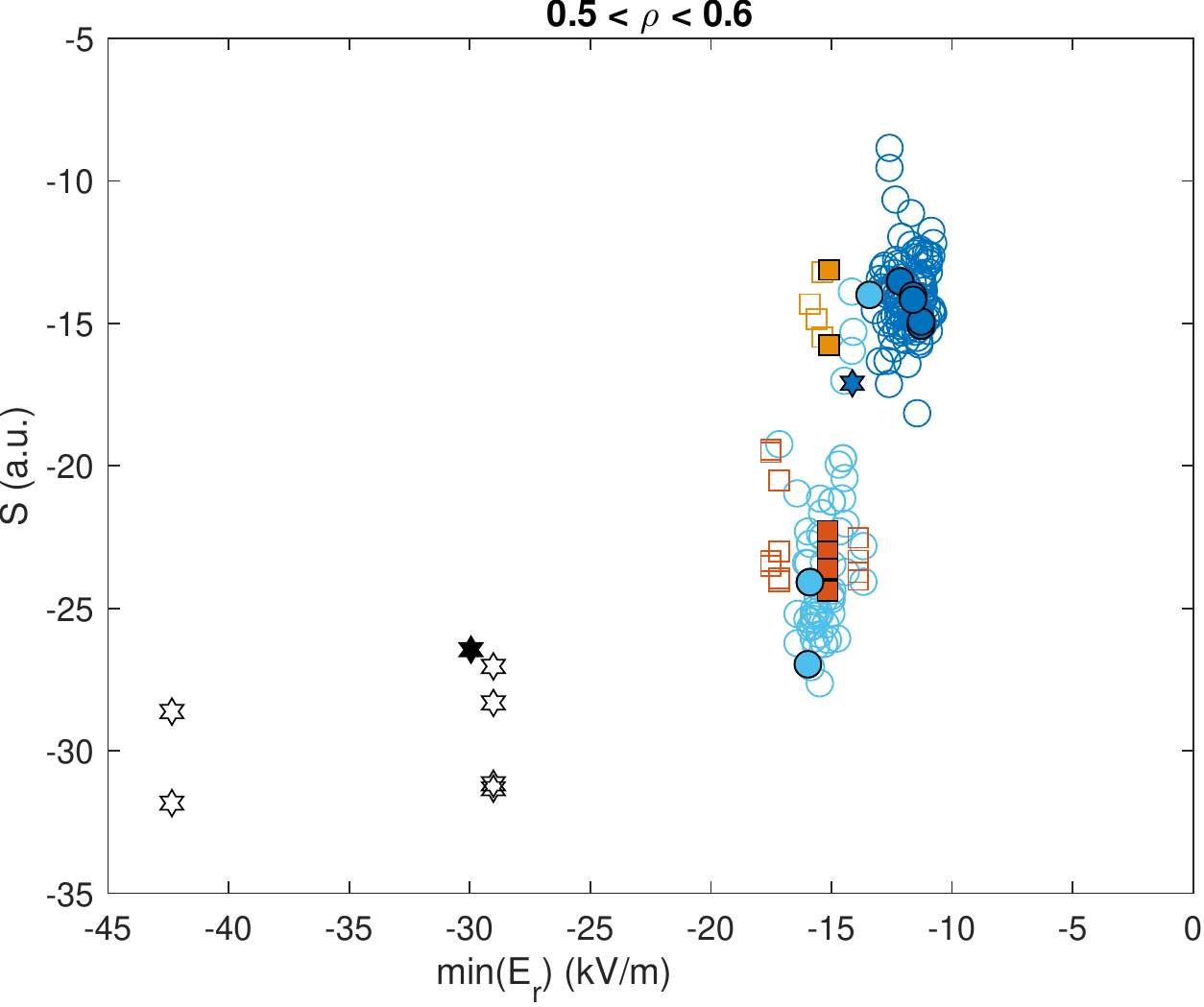}
	\caption{\textit{Fluctuation amplitude as a function of the local/minimum radial electric field (Top/bottom plot). Symbols and colors represent discharge families and representative discharges following previous convention.}}
	\label{fig4_d}
\end{figure}

Besides the value of local gradients, there are at least two other factors potentially influencing the growth of turbulent modes. The first of them is the local $T_e/T_i$ ratio, which is expected to destabilize ITG turbulence. In figure \ref{fig4_c} the measured fluctuation level is represented as a function of this ratio and no clear trends can be extracted from the data: as can be seen in figure \ref{fig2}, $T_e$ never takes values much higher than $T_i$ in the analyzed radial range (although it does for lower values of $\rho$). As a result $T_e/T_i < 1.5$ for the whole data set, and similar values of $T_e/T_i \simeq 1$ can be observed in the rather different high density ECRH and HP phases. Moreover, in the case in which a strongest change of this ratio is observed -low density ECRH discharges-, its increase of up to $50\%$ with respect to high density ECRH discharges corresponds to a strong decrease of fluctuations. This is contrary to expectations based on ITG models and reinforces the idea of some different physical phenomenon being responsible of the turbulence suppression observed in this family of discharges. The radial electric field is a second factor potentially influencing fluctuations, which are represented as a function of the former in figure \ref{fig4_d}. In the top plot, results already shown in figures \ref{fig3a} and \ref{fig3b} are summarized representing $S$ as a function of the local value of $E_r$ (i.e., values at the measurement location, without $|\nabla r|$ normalization to account for flux expansion in the magnetic surface \cite{Carralero20}): for most cases, $E_r$ values remain mostly unrelated to the amplitude of fluctuations, with $S$ changing over one order of magnitude while $E_r$ remains contained in a relatively narrow value range of $-15$ to $-10$ kV/m. Instead, substantially stronger electric fields are measured in the HP discharges, which also feature lower fluctuation amplitudes. In the bottom plot, fluctuations are represented as a function of the minimum value of the $E_r$ profile, which can be seen as a proxy for the corresponding general shearing effect of $E_r$. Although the effect is still weak for all but the HP scenarios, in this case a somewhat clearer trend can be observed, in which fluctuations tend to be reduced as the $E_r$ well deepens. This result is consistent with previous work \cite{Estrada21}, in which this was analyzed in detail by means of global linear gyrokinetic simulations using the code EUTERPE and it was concluded that, while $E_r$ and $E_r$ shear cause a reduction in the growth rate of instabilities, this contribution is small when compared to the effect of local gradients. Previous simulations carried out with the code GENE also found that, while $E_r$ causes a displacement of fluctuations towards regions of the magnetic surface with lower curvature thus contributing to the stabilization of the turbulence in W7-X, this effect is still secondary to the influence of profiles \cite{Xanthopoulos20}.\\

\section{Discussion and conclusions}\label{discussion}

The parametric dependencies of density fluctuations described in the previous chapter allow us to define a "baseline scenario", which is found under most operational regimes, including all ECRH heated plasmas beyond a certain density $\bar{n}_0 \simeq 5\cdot 10^{19}$ m$^{-3}$ and also many NBI-heated discharges (including in particular the discharge phases discussed earlier in which NBI power is combined with high levels of ECRH). In it, fluctuations show only a shallow decline towards the core, still having comparatively large amplitudes in the range of $S \simeq -15$ dB at the $0.5 < \rho < 0.6$ region. These discharges are characterized at the core by a relatively flat density profile with rather low $a/L_{n}$ values and core ion temperatures never exceeding the clamping  $T_{i,core} \simeq 1.7$ keV, thus featuring also moderate $a/L_{T_i}$ values. However, since the former is flatter than the latter, this leads to rather high values of $\eta_i \simeq 5-10$, which would be consistent with fully developed ITG modes. From this baseline scenario, two different pathways for core turbulence reduction have been observed in the surveyed operational regimes, one of which seems to be consistently related to changes in local values of gradients, while the second does not so clearly.\\

\begin{figure}
	\centering	
	\includegraphics[width=0.5\linewidth]{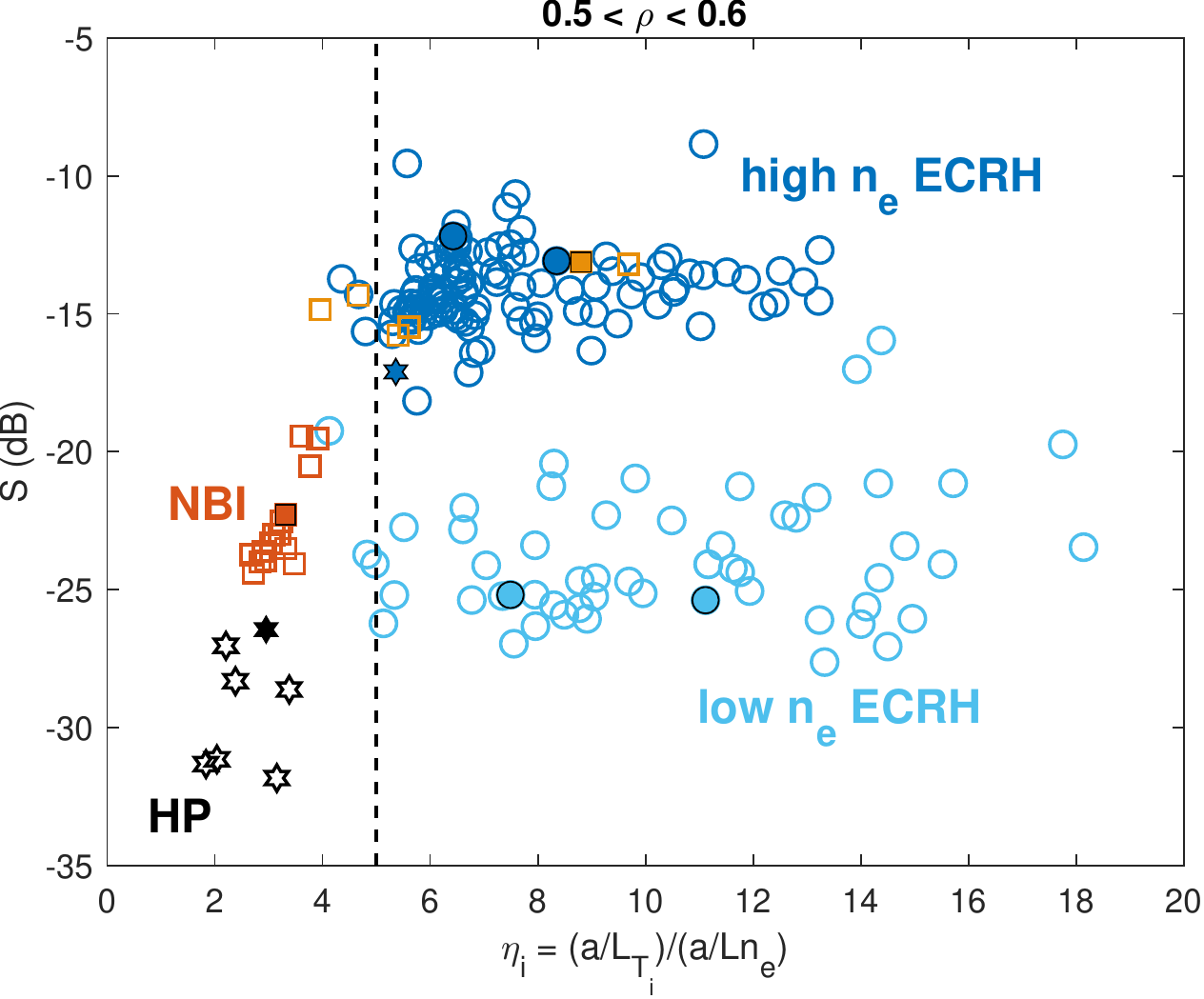}
	\caption{\textit{Fluctuations as a function of the gradient ratio $\eta_i$. Symbols and colors represent discharge families and representative discharges following previous convention. Vertical dashed line indicates the $\eta_i \simeq 5$ threshold discussed in the text.}}
	\label{fig5}
\end{figure}

The first pathway is represented by the third phase of NBI shots (featuring reduced ECRH heating, as seen in figure \ref{fig02}) and HP phases. In both cases, fluctuations are suppressed as density and ion temperature gradients increase. As shown in figure \ref{fig4_b}, the former increases faster than the latter in both cases, leading to a net reduction of $\eta_i$. If fluctuations are represented as a function of $\eta_i$, as is done in figure \ref{fig5}, this effect is manifest: a clear threshold is set between baseline discharges, which have a constant, high level of fluctuations for $\eta_i > 5$, while $S$ falls with $\eta_i$ for $\eta_i < 5$. Again, this behaviour of turbulence would be consistent with that of ITG modes, and the observed reduction in the scattered power measured by the DR would indicate lower turbulence amplitude as the result of the stabilization of such mode following the drop of $\eta_i$, in good agreement with theoretical expectations. According to these data, ITG  modes would be destabilized for some value $\eta_{i,c} \leq 2$ out of the measured range, and would saturate for values in the order of $\eta_i \simeq 5$, after which the baseline scenario sets in. Interestingly, this is consistent with linear gyrokinetic simulations carried out for W7-X, yielding an $\eta_{i,c} \simeq 1$ ITG critical threshold when the observed values of $T_e/T_i \simeq 1$ are considered \cite{Zocco18}. In the particular case of the HP discharges, the question remains about whether the strong observed $E_r$  may be also playing a role in the suppression of turbulence. In this sense, it must be noted that NBI and HP discharges seem to follow a common relation with $\eta_i$ despite their rather different $E_r$ values (in particular, it seems clear that $E_r$ does not play a role in the former), in line with the idea of $E_r$ playing a supporting but non-essential role in turbulence suppression as already discussed in the previous section. A word of caution is in order, though: while the different groups seem to be sufficiently separated and the existence of some threshold between them is clear, the quality of the fits to the experimental profiles of density and temperature is probably not enough to attempt a quantitative description. Therefore, exact values such as the $\eta_i = 5$ limit or the apparent value of the $S$ vs. $\eta_i$ slope should be approached with care.\\

The second pathway is represented by low density ECRH discharges. In this case, a reduction of fluctuations is observed  without a clear change in the local gradients. In particular, as can be seen in figure \ref{fig5}, these points feature the same or even higher values of $\eta_i$ than those of the high density discharges, and never go below the previously introduced threshold at $\eta_i \simeq 5$. As discussed in the previous section and shown in figures \ref{fig3b} and \ref{fig4_d}, no major changes are observed either in $E_r$ which could justify this reduction, specially in the radial region under discussion. Looking at figure \ref{fig2} it can be seen that the main difference between the two groups of ECRH discharges at the core is that those with lower densities also feature substantially larger $T_e$ values. As a consequence of this, the $T_e/T_i$ ratio is also somewhat increased in them as shown in figure \ref{fig4_c}, which makes once again the reduction of fluctuations inconsistent with the expectations for ITG turbulence.  Another consequence of the lower densities and higher electron temperatures is that electron collisionality, $\nu_{e} \propto n_eT_e^{-3/2}$ is substantially lower in this group of discharges. This points towards another possible explanation for the observed suppression of turbulence: since a strong $a/L_{T_e}$ is present in all the ECRH scenarios (specially so in the low density ECRH discharges), a reduction of collisionality below a certain threshold might destabilize some TEM which would then become the dominant turbulent mode at the expense of the ITG. As discussed in the Introduction, such process has been reported in tokamaks and supported by gyrokinetic GS2 simulations \cite{Ryter05}. In order to assess the stabilizing effect of collisionality on TEMs in the described discharges, a normalized electron frequency $\nu_e^* := \nu_{eff} \tau_{o}$ can be defined, where $\nu_{eff} := \nu_{e} R/a$ is the effective collision frequency of trapped electrons, $\tau_{o}^{-1} \simeq \epsilon^{1/2} v_{th,e}/R$ is the typical trapped electron orbit time, $v_{th,e}$ is the thermal velocity of electrons and $\epsilon = a/R$ is the inverse aspect ratio. If $\nu_e^* \ll 1$, trapped electrons may complete their orbits unimpeded by collisions thus allowing the onset of TEM turbulence, which is suppressed by collisionality otherwise.\\

\begin{figure}
	\centering	
	\includegraphics[width=0.5\linewidth]{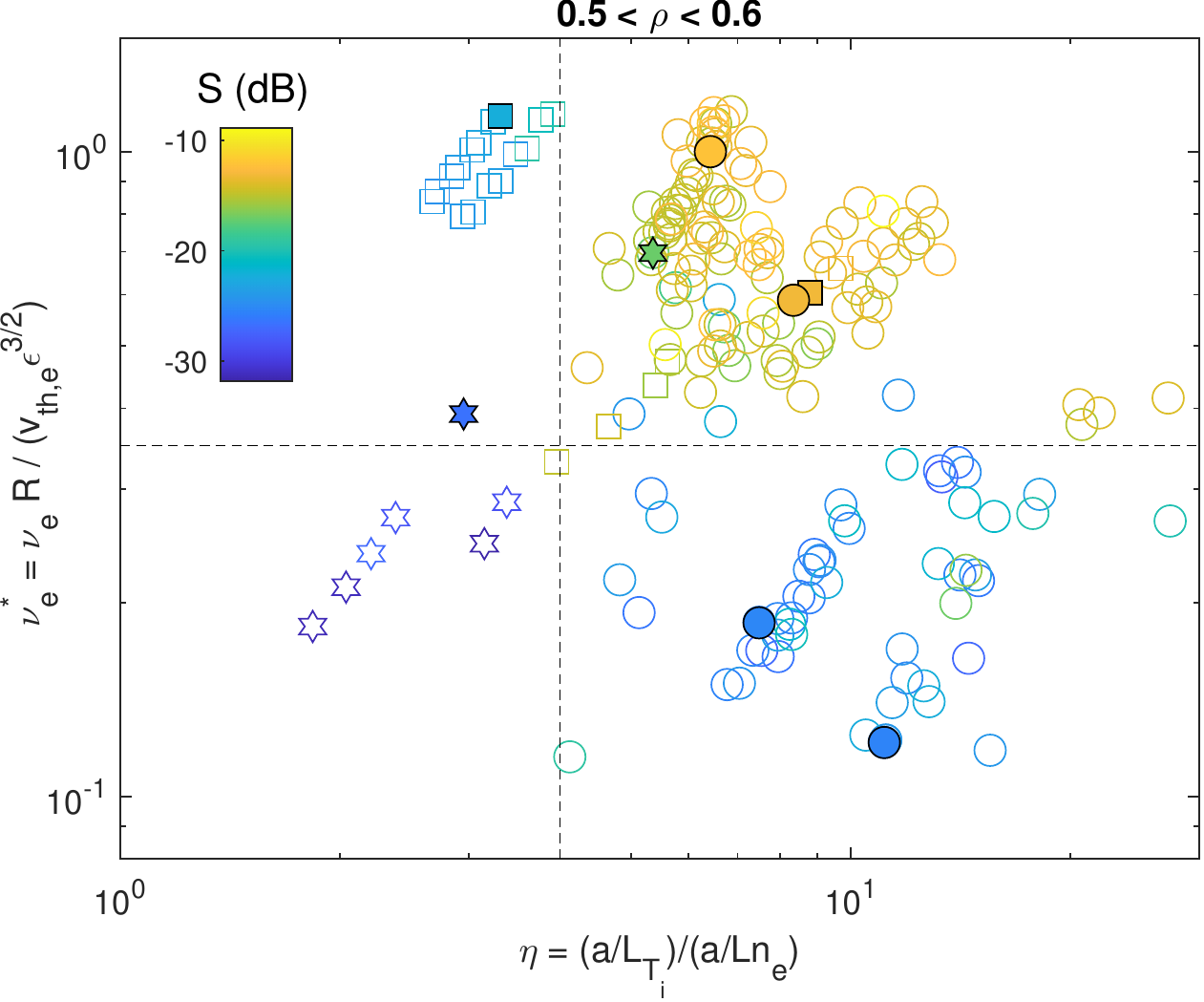}
	\caption{\textit{Fluctuation amplitude (in color) as a function of gradient ratio $\eta_i$ and collisionality ratio $\nu^*_{e}$. Symbols represent discharge families following previous convention, with solid ones indicating representative discharges. Dashed lines represent the thresholds $\eta_i \simeq 5$ and $\nu^*_{e} \simeq 0.35$, discussed in the text. }}
	\label{fig6}
\end{figure}

In order to discuss this effect, fluctuations are represented in figure \ref{fig6} as a function of $\eta_i$ and $\nu^*_{e}$. As can be seen, fluctuation amplitude is now well classified in four quarters, separated by two threshold vales. The first is the $\eta_i \simeq 5$ threshold already discussed in figure \ref{fig5}, and separates third phase NBI and HP points from the rest of the database. However, when considering ECRH discharges, a very clear second threshold appears at $\nu^*_{e} \simeq 0.35$, which effectively separates high density ECRH discharges, featuring strong fluctuations, and low density ECRH ones, for which turbulence is suppressed. This sharp separation of the two groups of points, along with the reasonable value of the threshold and the already discussed different behaviour of the low ECRH discharges, is a strong indication that this reduction might be caused by a transition from ITG-dominated to TEM-dominated turbulence. The general implications of such transition would be rather complex and are out of the scope of the present work. However, it could have some effects which would lead to a clear reduction in the fluctuations, as observed by the DR (although, not necessarily to a global reduction of turbulence or turbulent transport): 
In particular, the region in real space with the most unstable modes would move away from the outer midplane of the bean section being probed by the DR -and where ITG modes are destabilized-, to regions with higher trapped electron fractions, which are generally away from the measurement zone \cite{Proll13}. In any case, while this hypothesis seems reasonable and could explain the observed phenomena, either further empirical evidence or realistic simulations of the two regimes carried out with some gyrokinetic code including kinetic electrons would be required to confirm it. This is left for future work.\\

\begin{figure}
	\centering	
	\includegraphics[width=0.5\linewidth]{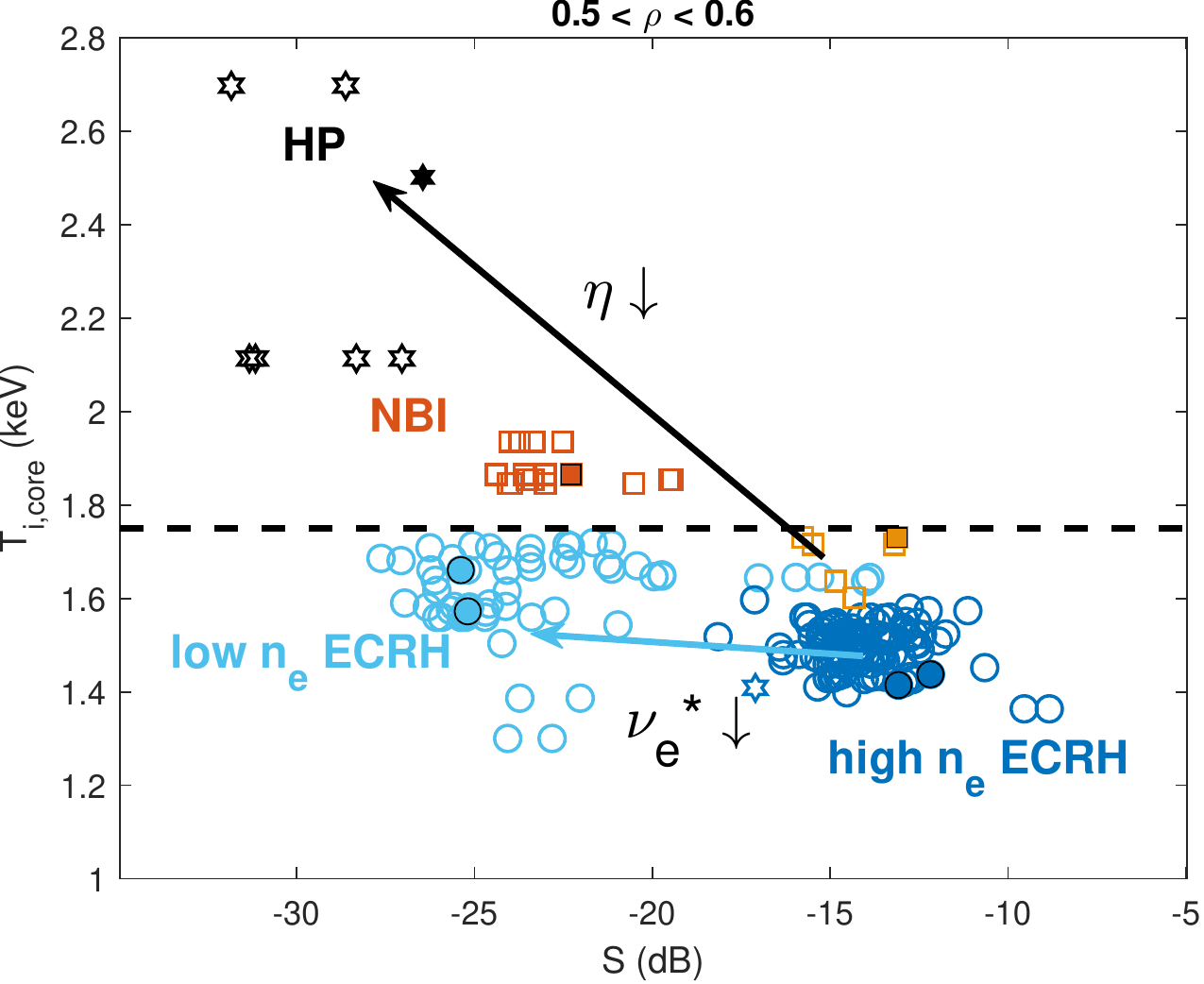}
	\caption{\textit{Core ion temperature as a function of fluctuations at the measurement region $0.5 < \rho < 0.6$. Symbols and colors represent discharge families and representative discharges following previous convention. Dashed line represents the  $T_{i,core} \simeq 1.7$ limit discussed in the Introduction. Black and blue arrows indicate the two pathways to turbulence reduction, respectively as a result of the decline of $\eta_i$ and $\nu_{e}$.}}
	\label{fig7}
\end{figure}

Regarding the relation between turbulence suppression and the global performance of W7-X, some preliminary observations can already be made from the data discussed so far. In particular, it can be seen how turbulence suppression is observed in all regimes in which the $T_{i,core} \simeq 1.7$ limit is exceeded: in figure \ref{fig7}, $T_{i,core}$ is represented as a function of the turbulence levels in the analyzed region $0.5 < \rho < 0.6$, and the same trend discussed in figure \ref{fig5} can be seen in it, going from baseline cases into third phase NBI ones (which go slightly over the threshold) and eventually into the HP shots, which clearly overpass the clamping. This result, which could be extended to less frequent situations in which this limit is surpassed such as the deliberate injection of Boron \cite{Lunsford21}, suggests that the local drop of turbulence observed by the DR when $\eta_i$ is reduced is representative of a global suppression of turbulent transport, leading to an enhanced performance of the machine. Finally, the behaviour of the $\eta_i$-suppression is in contrast with that of the $\nu_{e}$-suppression: low density ECRH discharges also display a strong suppression of turbulence, but it does not lead to high $T_{i,core}$ values, supporting the hypothesis of a change in the kind of turbulence rather than a global suppression. This picture is however somewhat complicated by the transient nature of the $T_{i,core}$ improvement in the NBI discharges: the continued increase of density observed in that example leads to the reversal to $T_{i,core} \leq 1.7$ keV  below the clamping limit while $a/L_n$ increases, keeping low values of $\eta_i$ and, in good agreement with the ITG suppression hypothesis, consequently also low values of fluctuations. This way, the stabilization of the ITG and the subsequent reduction of turbulent transport would be a necessary condition for the achievement of high ion temperatures at the core, but might require additional conditions related to the heating and fueling conditions (which, as explained earlier, are substantially different in NBI scenarios). In order to clarify this point, the relation between local fluctuations and global turbulent transport will need to be systematically assessed for the discussed scenarios. This task will be addressed in a forthcoming work. \\

In conclusion, we have created a database of DR measurements of ion-scale turbulence from the core of W7-X, representative of a number of relevant scenarios carried out in the standard configuration during the last experimental campaign. These measurements have then been cross-referenced with local measurements of gradients and other parameters potentially influencing the stability of turbulent modes, such as the $T_e/T_i$ ratio, $E_r$ and its shear. By doing so, we have identified a number of different scenarios with respect to core turbulence amplitude: first, there is common scenario (thus designated as "baseline") found under most ECRH gas puffing and many NBI discharges, in which both fluctuation amplitude and $\eta_i$ are high, and $T_{i,core} \leq 1.7$ keV. Therefore, and taking into account that fluctuation measurements come from the region of bad curvature, this is considered consistent with ITG-dominated turbulence. Also in good agreement with such conclusion, amplitude of density fluctuations is observed to drop clearly as $\eta_i \lesssim 5$ in the high performance phase (ECRH heating after pellets) and one NBI scenario featuring limited ECRH heating. In this case, temperature ratio and $E_r$ seem to play a secondary role, but behave according to expectations. Other family of discharges also show a strong reduction of fluctuations when density is reduced in ECRH plasmas. In this case, neither $\eta_i$ nor $E_r$ seem to change substantially, and $T_e/T_i$ is actually increased (which should lead to a further destabilization of an ITG mode). Instead, it is found that this suppression is related to a $\nu^*_{e} \simeq 0.35$ collisionality threshold, therefore suggesting a possible ITG to TEM transition changing the region in which modes destabilize preferentially and moving it away from the DR measurement zone. Finally, it can be stated that for all scenarios in which the usual $T_{i,core} \simeq 1.7$ keV is exceeded, a reduction of turbulence is observed in the $0.5 < \rho < 0.6$ region following a clear reduction in the local value of $\eta_{i}$, experimentally validating the idea that the temperature clamping is associated to turbulent transport and pointing towards the suppression of ITG turbulence as a necessary condition mechanism for it. This will be studied at length in a forthcoming work which will deal with the relation between observed fluctuations and turbulent transport, thus trying to provide a solid link between different scenarios of operation in W7-X and microturbulence behaviour, and contributing to the design of future experiments in which transport can be kept under control and projected core temperatures eventually achieved.\\

\section*{Acknowledgments}

The authors acknowledge the entire W7-X team for their support. This work has been partially funded by the Spanish Ministry of Science and Innovation under contract number FIS2017-88892-P and partially supported by grant ENE2015-70142-P, Ministerio de Economía y Competitividad, Spain and by grant PGC2018-095307-B-I00, Ministerio de Ciencia, Innovación y Universidades, Spain. This work has been sponsored in part by the Comunidad de Madrid under projects 2017-T1/AMB-5625 and Y2018/NMT [PROMETEO-CM]. This work has been carried out within the framework of the EUROfusion Consortium and has received funding from the Euratom research and training programme 2014-2018 and 2019-2020 under grant agreement No 633053. The views and opinions expressed herein do not necessarily reflect those of the European Commission.\\

\end{document}